\documentclass[11pt,a4paper]{article} 
\pdfoutput=1
\usepackage{jheppub} 
\usepackage{graphicx} 
\usepackage{amsmath} 
\usepackage{amssymb} 
\usepackage{slashed} 
\usepackage{bigstrut} 
\usepackage{mathtools} 
\usepackage{multirow} 
\usepackage[utf8]{inputenc}
\usepackage{lineno}
\usepackage{todonotes}
\usepackage{xspace}
\usepackage{lscape}
\usepackage{wasysym}
\usepackage[calc,useregional]{datetime2}

\newcommand{\msbar}{\ensuremath{\overline{\rm MS}}\xspace}

\newcommand{\mtpole}{\ensuremath{m_{t}^{\text{pole}}}\xspace}
\newcommand{\ttbar}{\ensuremath{t\overline{t}}\xspace}
\newcommand{\mtt}{\ensuremath{M(\ttbar)}\xspace}
\newcommand{\ytt}{\ensuremath{|y(\ttbar)|}\xspace}
\newcommand{\chisq}{\ensuremath{\chi^{2}}\xspace}

\title{

  Top-quark pole mass extraction at NNLO accuracy, from total, single- and double-differential cross sections for $t\bar{t}+X$ production at the LHC
}

\author[a]{M.V.~Garzelli,}
\author[b]{~J.~Mazzitelli,}
\author[a]{~S.-O.~Moch,} 
\author[a]{~O.~Zenaiev}

\affiliation[a]{II.~Institute for Theoretical Physics, Hamburg University \\
Luruper Chaussee 149, D--22761 Hamburg, Germany}
\affiliation[b]{Paul Scherrer Institut, CH--5232 Villigen, Switzerland}

\emailAdd{maria.vittoria.garzelli@desy.de} 
\emailAdd{javier.mazzitelli@psi.ch} 
\emailAdd{sven-olaf.moch@desy.de} 
\emailAdd{oleksandr.zenaiev@desy.de}

\abstract{
  We extract the top-quark mass value in the on-shell renormalization scheme from the comparison of theoretical predictions for $pp \rightarrow t\bar{t} + X$ at next-to-next-to-leading order (NNLO) QCD accuracy with experimental data collected by the ATLAS and CMS collaborations for absolute total, normalized single-differential and double-differential cross-sections during Run~1, Run~2 and the ongoing Run~3 at the Large Hadron Collider (LHC). For the theory computations of heavy-quark pair-production we use the \texttt{MATRIX} framework, interfaced to \texttt{PineAPPL} for the generation of grids of theory predictions, which can be efficiently used a-posteriori during the fit, performed within \texttt{xFitter}. We take several state-of-the-art parton distribution functions (PDFs) as input for the fit and evaluate their associated uncertainties, as well as the uncertainties arising from renormalization and factorization scale variation. Fit uncertainties related to the datasets are also part of the extracted uncertainty of the top-quark mass and turn out to be of similar size as the combined scale and PDF uncertainty. Fit results from different PDF sets agree among each other within 1$\sigma$ uncertainty, whereas some datasets related to $t\bar{t}$ decay in different channels (dileptonic vs. semileptonic) point towards top-quark mass values in slight tension among each other, although still compatible within $2.5\,\sigma$ accuracy. Our results are compatible with the PDG 2022 top-quark pole-mass value. Our work opens the road towards more complex simultaneous NNLO fits of PDFs, the strong coupling $\alpha_s(M_Z)$ and the top-quark mass, using the currently most precise experimental data on $t\bar{t} + X$ total and multi-differential cross sections from the LHC.
  }


\preprint{DESY-23-179, PSI-PR-23-42}
\arxivnumber{arXiv:2311.05509} 
\begin{document} 
\maketitle

\section{Introduction}
\label{intro}

Top-quark measurements at the Large Hadron Collider (LHC) play a pivotal role in modern particle physics for a number of reasons. First, they are crucial for precisely extracting key parameters of the Standard Model (SM), helping to refine our general understanding of fundamental interactions. Second, these measurements provide critical insights into the electroweak symmetry breaking mechanism, shedding light on how particles acquire mass. Finally, top-quark studies are a vital component of searches for physics beyond the SM (BSM) as one of the most important backgrounds, but also for potentially uncovering new phenomena, e.g.\ through anomalous couplings to top quarks. In this context, measurements of $t\bar{t} + X$ hadroproduction serve as a cornerstone of the LHC physics program.

The significance of top-quark physics in ongoing and forthcoming research at high-energy colliders has been acknowledged, see e.g.\ Refs.~\cite{Cristinziani:2016vif,Husemann:2017eka,Chivukula:2017vza}, and the need for accurate theoretical predictions accompanying experimental studies has been underlined and motivated again in Ref.~\cite{Schwienhorst:2022yqu} summarizing the recent Snowmass 2021 process.
The present study aims at the determination of the top-quark mass, a topic 
 which has been extensively reviewed, e.g., in Refs.~\cite{Moch:2014tta, Beneke:2016cbu, Nason:2017tpm, Hoang:2020iah}.
The pole mass of the top quark is extracted from a com\-pa\-ri\-son of inclusive and differential cross-section data for $t\bar{t}+X$ production collected 
by the LHC experiments ATLAS and CMS, 
 with theoretical predictions including
 higher-order corrections in quantum chromodynamics (QCD) and computed for the top-quark mass in the on-shell renormalization scheme.

From the theory point of view, predictions for $pp \rightarrow t\bar{t} + X$
at next-to-leading order (NLO) QCD accuracy were presented already many years ago, starting from the works
of Refs.~\cite{Nason:1987xz, Nason:1989zy, Beenakker:1990maa}.
Total cross-sections at next-to-next-to-leading order (NNLO) accuracy
are available already since more than ten years~\cite{Czakon:2013goa}. 
Public codes like, e.g., {\texttt{HATHOR}}~\cite{Aliev:2010zk} and {\texttt{Top++}}~\cite{Czakon:2011xx} have opened the possibility to access to them already since long. Predictions first at approximate NNLO (aNNLO)~\cite{Moch:2008qy, Guzzi:2014wia} 
and then at approximate N$^3$LO (aN$^3$LO)~\cite{Kidonakis:2023juy} accuracy obtained from fixed-order expansions of threshold-resummed results at next-to-next-to-leading logarithmic accuracy, have been also produced, targeting both total and differential cross-sections.

Limiting the discussion to fixed-order calculations, NNLO QCD differential cross sections have been computed more recently than NNLO total cross sections. First computations in this direction were performed in Ref.~\cite{Czakon:2016ckf} for $p\bar{p}$ collisions and in Refs.~\cite{Czakon:2015owf, Czakon:2016dgf} for $pp$ collisions, using the sector-improved residue subtraction scheme {\texttt{STRIPPER}}~\cite{Czakon:2014oma, Czakon:2011ve, Czakon:2010td} 
for the cancellation of infrared divergences at NNLO, born from the combination of some ideas of FKS NLO subtraction~\cite{Frixione:1995ms} with those of sector decomposition~\cite{Binoth:2004jv, Anastasiou:2003gr, Binoth:2000ps}.
Although the corresponding code is still private, part of the results of these computations are nowadays accessible to the whole HEP community through the \texttt{HighTEA} analyzer project~\cite{Czakon:2023hls}, following a first release as fastNLO~\cite{Britzger:2012bs} grids~\cite{Czakon:2017dip}. In parallel, partial NNLO results, limited to the $q\bar{q}$ channel~\cite{Abelof:2014fza, Abelof:2014jna, Abelof:2015lna}, were also obtained by a group developing and using the antenna subtraction method~\cite{Gehrmann-DeRidder:2005btv}, properly extended to deal with colorful initial states and colorful massive final states~\cite{Abelof:2011jv}.
More recently, the $q_T$-subtraction method~\cite{Catani:2007vq} was also extended and applied to the calculation of the cross sections for this process, starting from the off-diagonal channels in Ref.~\cite{Bonciani:2015sha} and later presenting complete NNLO results for the total and differential cross sections in Refs.~\cite{Catani:2019iny} and \cite{Catani:2019hip}, respectively.
The results were obtained and implemented within the \texttt{MATRIX} framework~\cite{Grazzini:2017mhc}. 
Those implementations adopt the on-shell top-quark mass renormalization scheme. As an alternative, following \cite{Langenfeld:2009wd}, in Ref.~\cite{Catani:2020tko} predictions with top-quark mass renormalized in the ${\overline{\mathrm{MS}}}$ scheme were presented.
The implementation using the on-shell top-quark mass renormalization scheme has been made available in the public version of \texttt{MATRIX} that, in principle, enables the whole HEP community to make predictions of single- and even double-differential cross sections by just installing and running the code after having specified some inputs. In this way it is possible to obtain predictions which can be directly compared to the experimental data released during Run~1, 2 and 3 at the LHC (see e.g. the experimental data published in Refs.~\cite{ta5, tc5, tca78, ta136, tc136, top20001, tc13ll, ta13lj, ta13ll, top18004,a190807305,a200609274,top14013,a14070371,a151104716,a160707281}). 

In practice, the amount of computations for different sets of input parameters required to calculate the uncertainties on these predictions, in particular due to parton distribution functions (PDF) and the top-quark mass, 
is quite large and very demanding in terms of computing resources. Strategies have been proposed and/or already developed to improve the speed of these computations, shortening the processor and memory usage. Saving the results in grids, which can be used a-posteriori via interpolation for further analyses, e.g. fits of PDFs and/or top-quark mass value, turns out to be an indispensable step, at least considering present computing resources. In Ref.~\cite{Guzzi:2014wia} a PDF determination was carried out using predictions at aNNLO accuracy computed with \texttt{DiffTop} interfaced to \texttt{fastNLO}~\cite{Britzger:2012bs} and \texttt{xFitter}~\cite{Alekhin:2014irh}. The importance of saving NNLO predictions in grids was discussed in Ref.~\cite{Czakon:2017dip}, and a PDF fit using them for single-differential \ttbar distributions was presented in Ref.~\cite{Czakon:2016olj}, a study where NNLO results were mimicked by using NLO calculations interfaced to \texttt{APPLgrid}~\cite{Carli:2010rw} and supplementing them with bin-by-bin $K$-factors. The latter were defined as the ratio of the NNLO to NLO predictions for a single PDF set, and the same $K$-factors were even applied to NLO predictions obtained with different PDFs, thus ignoring their implicit dependence of the PDFs. A simultaneous fit of the top-quark mass and the strong coupling $\alpha_s(M_Z)$ at NNLO using single-differential $\ttbar+X$ distributions was performed in Ref.~\cite{Cooper-Sarkar:2020twv}. The scale dependence of the top-quark mass in the \msbar renormalization scheme, i.e. its running through NNLO in QCD, has been investigated in Refs.~\cite{CMS:2019jul,Defranchis:2022nqb} using data from the CMS collaboration for the single-differential cross section in the invariant mass of the \ttbar system.

In the current study, thanks to tools that we describe in more detail in Section~\ref{sec:matrixtheory}, it was possible for us to use a customized version of \texttt{MATRIX}, optimized for the $pp \rightarrow t\bar{t} + X$ process and interfaced to \texttt{PineAPPL}~\cite{Carrazza:2020gss}, for the computation of all NNLO QCD theory predictions with uncertainties (without utilizing $K$-factors or other approximations) for absolute total as well as normalized single- and double-differential cross sections at the LHC, that we have included in our fits of top-quark pole mass values at NNLO accuracy. 
We use experimental measurements of cross sections as a function of the invariant mass \mtt and rapidity \ytt of the \ttbar pair.
These are the first fits of the top-quark pole mass with NNLO accuracy using as input LHC double-differential $t\bar{t} + X$ data, to the best of our knowledge. 
For completeness, it is worth to also notice that some of the double-differential data considered in our work have been used in the extraction of PDFs, comparing them to theoretical predictions obtained using NNLO/NLO K-factors (see e.g. Refs.~\cite{Ablat:2023tiy,NNPDF:2021njg}).
Furthermore, the impact of the CMS double-differential production cross-section dataset at 8 TeV~\cite{top14013} on the gluon PDFs has been investigated in Ref.~\cite{Czakon:2019yrx}
by comparing them to genuine NNLO double-differential predictions
in fastNLO tables obtained within the STRIPPER approach.

A comparison of our predictions before fit, to the experimental data from the ATLAS and CMS collaborations considered in the fit is reported in Section~\ref{sec:compa}, whereas the fit methodology and results are presented in Section~\ref{sec:fit}. 
A summary and perspectives for future developments are presented in the Conclusions in Section~\ref{sec:conclu}.


\section{Strategy for high-performance computations of NNLO (double-)differential cross-sections for $t\bar{t} + X$ production at the LHC} 
\label{sec:matrixtheory}

\subsection{$q_T$ subtraction and the \texttt{MATRIX} framework}

The $q_T$-subtraction formalism is a method for handling and cancelling the infrared divergences from the combination of real and virtual contributions in computations of total and differential cross sections for the production of massive final states in QCD at NLO and NNLO accuracy.  
It uses as a basis the $q_T$-resummation formalism, where $q_T$ is the transverse momentum of the produced high-mass system. The latter allows to compute the infrared subtraction coun\-ter\-terms, constructed by the evaluation of the $q_T$ distribution of the massive final-state system in the limit $q_T \rightarrow 0$. 
In the case of a colorless massive final state, the $q_T$ distribution in this limit has a universal structure, known at N$^3$LO accuracy from the expansion of the corresponding resummed result. 
 $q_T$ resummation, and, as a consequence, $q_T$ subtraction, were first developed and applied to the inclusive hadro-production of Higgs~\cite{Catani:2007vq} and vector~\cite{Catani:2009sm} bosons, and then extended to the case of heavy-quark pair hadro-production~\cite{Zhu:2012ts, Li:2013mia, Catani:2014qha, Catani:2018mei}. In the case of final states containing heavy quarks, which are colored objects, additional soft singularities appear, absent in the case of colorless final states. No new collinear singularity enters the calculation, thanks to the heavy-quark mass. Recently, the methodology was used even for first NNLO computations of heavy-quark pair production in association with a Higgs~\cite{Catani:2022mfv} or a $W$-boson~\cite{Buonocore:2022pqq, Buonocore:2023ljm} at the LHC.

The master formula for the differential (N)NLO partonic cross section for $pp\rightarrow t\bar{t}+X$ production following this formalism, can be written as 
\begin{equation}
\label{eq:Xsection}
d\sigma^{t\bar{t}}_{(N)NLO} = \mathcal{H}^{t\bar{t}}_{(N)NLO} \otimes d\sigma^{t\bar{t}}_{LO}
+ \left[ d\sigma^{t\bar{t}+jet}_{(N)LO} - d\sigma^{t\bar{t},CT}_{(N)NLO} \right]
\end{equation}
where $d\sigma^{t\bar{t}}_{LO}$ is the differential cross section at LO accuracy, $d\sigma^{t\bar{t}+jet}_{(N)LO}$ is the cross section for $t\bar{t}j + X$ production at (N)LO accuracy, built according to the standard LO and NLO formalisms and  $d\sigma^{t\bar{t},CT}_{(N)NLO}$ are appropriate counterterms capturing the singular behaviour of $d\sigma^{t\bar{t}+jet}_{(N)LO}$ in the limit $q_T \rightarrow 0$, such that the content of the square bracket is infrared-safe in the limit $q_T \rightarrow 0$.
These counterterms are built from the (N)NLO 
fixed-order expansion of the $q_T$ resummation formula for $t\bar{t}$ production (see also Ref.~\cite{Ju:2022wia}).

The form of these counterterms is nowadays fully known. 
In practice, the computation of the part in square brackets in eq.~(\ref{eq:Xsection}) is carried out 
introducing a cut-off $r_0 = q_{T, min}/\mtt$ on the dimensionless quantity $r = q_T/\mtt$, where \mtt is the invariant mass of the $t\bar{t}$ pair.
This renders both terms in the square brackets separately finite. 
Below $r_0$, which acts effectively as a slicing parameter, the two terms in square brackets are assumed to coincide, which is correct up to power-suppressed contributions. 
These power corrections vanish in the limit $r_0 \rightarrow 0$, and as a mitigation strategy, to control their impact on the cross section in eq.~(\ref{eq:Xsection}), \texttt{MATRIX} computes the quantity in square brackets for different discrete $r_0$ values in the same run. The results for different $r_0$ values are then fitted, and the extrapolation to $r_0=0$ is taken, to get information on the exact NNLO result. Extrapolation uncertainties are also accounted for and included, as explained in Ref.~\cite{Grazzini:2017mhc}. 
A phase-space slicing approach like the $q_T$-subtraction formalism can also be
subject to linear power corrections from fiducial cuts on the final state decay products (see studies for the Drell-Yan process in Refs.~\cite{Ebert:2020dfc,Alekhin:2021xcu}). So far, such effects have been treated in \texttt{MATRIX} only for color-singlet final states~\cite{Buonocore:2021tke}.

The term $\mathcal{H}^{t\bar{t}}_{(N)NLO}$ in eq.~(\ref{eq:Xsection}) includes information on the virtual corrections to the process at hand and contributions that compensate for the subtraction of the counterterms.  $\mathcal{H}^{t\bar{t}}_{(N)NLO}$ can be splitted into a process-independent and a process-dependent part. The process-independent part is the same entering Higgs- and vector-boson production, and is explicitly known~\cite{Catani:2011kr, Catani:2012qa, Gehrmann:2012ze, Gehrmann:2014yya}. 
The process-dependent part of $\mathcal{H}^{t\bar{t}}_{NNLO}$ in the flavour-off-diagonal channels 
(i.e., all except for $q\bar{q}$ and $gg$)
originates from one-loop amplitudes for the partonic processes   $q\bar{q} \rightarrow t\bar{t}$ and $gg \rightarrow t\bar{t}$, plus the heavy-quark azimuthal correlation terms~\cite{Catani:2014qha} in the $q_T$-resummation formalism. 
The process-dependent part of $\mathcal{H}^{t\bar{t}}_{NNLO}$  in the flavour-diagonal channels 
($q\bar{q}$ and $gg$)
requires in addition the knowledge of two-loop amplitudes for $q\bar{q} \rightarrow t\bar{t}$ and $gg \rightarrow t\bar{t}$ for which the results of Refs.~\cite{Czakon:2008zk, Barnreuther:2013qvf} are adopted, plus the computation of additional soft contributions, completed in Ref.~\cite{Catani:2023tby}. 
More details can be found in Refs.~\cite{Catani:2014qha,Bonciani:2015sha,Catani:2023tby}.
All these ingredients are available and have been combined together in the \texttt{MATRIX} framework, which, in turn, relies on the \texttt{MUNICH} code for the combination of real and virtual NLO contributions and for the evaluation of  $d\sigma_{(N)NLO}^{t\bar{t}, CT}$, on additional code for the evaluation of $\mathcal{H}^{t\bar{t}}_{(N)NLO}$, and on the \texttt{OpenLoops} code for the evaluation of all color-correlated and spin-correlated tree-level and one-loop amplitudes~\footnote{The four-parton color-correlations are treated separately by an analytical implementation.}. 

The \texttt{MATRIX} computer program has been kept quite general and this allowed the implementation of a number of different processes in a comprehensive framework. We use a customized version of \texttt{MATRIX}, tailored to the $t\bar{t}~+~X$ case only and optimized for it. In particular, we started from the \texttt{MATRIX} version used for the computations presented in Ref.~\cite{Catani:2019hip} and we have performed a number of optimizations in the program flow and execution, which include 1) recycling of parts of computations already done, instead of recomputing multiple times identical pieces, 2) adaptation of the code in view of its execution on local multicore machines~\footnote{The computations of this work have made use of the high-performance computing infrastructure provided by the DESY BIRD (Big-data Infrastructure for Research and Development) computing cluster.}, 3) optimizations in distributing the computation on different machines/cores, in the job and job failure handling, 4) optimization in the input/output information exchange with the computer  cluster during remote job execution, 
5) reduction in the memory usage and in the size of the stored output, leading to an overall gain in the speed of the computation, in the memory consumption and in the space allocation without compromising the final results. 
The exact gain in the speed, memory and disk space consumption depends on the required accuracy for the cross section to be computed and the number of parallel jobs. In our case, these modifications allowed us to reach the desired $0.2\permil$ accuracy for the total $\ttbar+X$ cross sections.

\smallskip
All our computations apply the on-shell renormalization scheme for the top-quark mass. 
The theoretical aspects, advantages and disadvantages of this choice are well known~\cite{Hoang:2020iah}.
In particular, cross sections for $t\bar{t} + X$ hadro-production can be subject to another class of power corrections 
that originate from renormalons,
see e.g., Ref.~\cite{Makarov:2023uet}.
Such effects or any estimates of their size are not included in our analysis.

Also, electroweak corrections at NLO accuracy~\cite{Kuhn:2006vh} are not included in our calculations. According to Ref.~\cite{Czakon:2017wor}, and considering the thin binning in that work, their size varies from $+2\%$ to $-4\%$ as a function of \mtt, and within $1\%$ as a function of \ytt. However, we have checked that for the bins of the experimental measurements which we use in our work this effect does not exceed~$1\%$. Thus, the missing electroweak corrections are expected to be covered by the assigned theoretical uncertainties described in Section~\ref{sec:pineappl}. 

\begin{figure}[htb]
    \centering
    \includegraphics[width=0.49\textwidth]{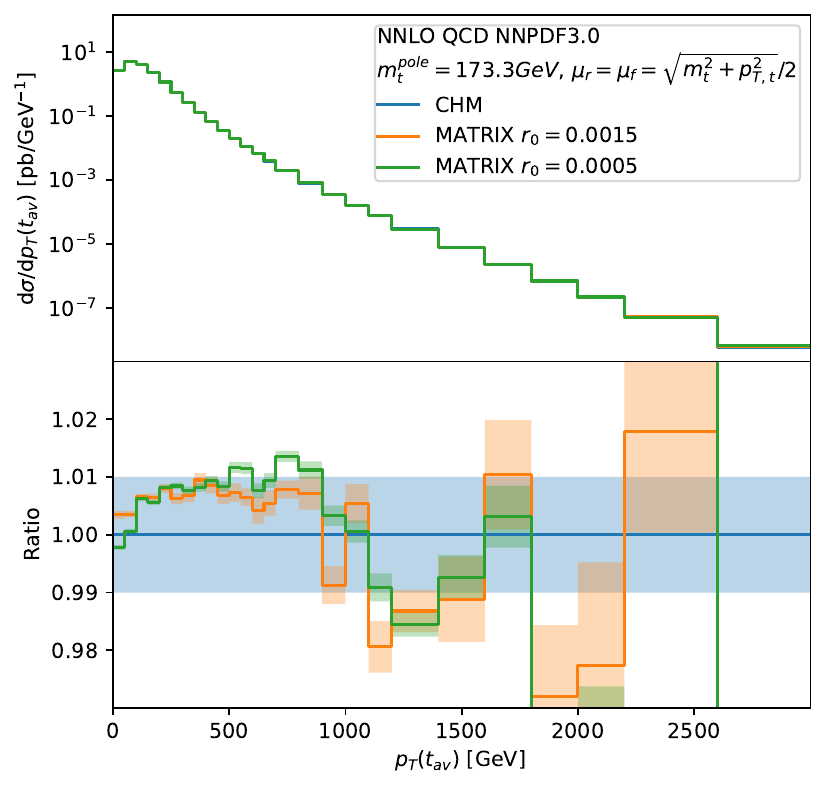}
    \includegraphics[width=0.49\textwidth]{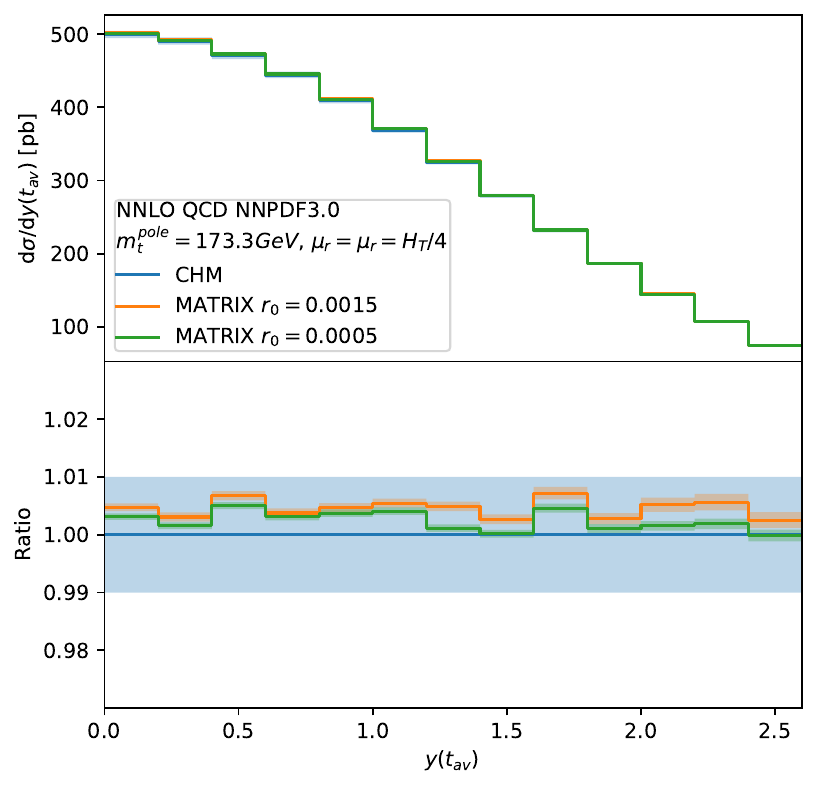}
    \includegraphics[width=0.49\textwidth]{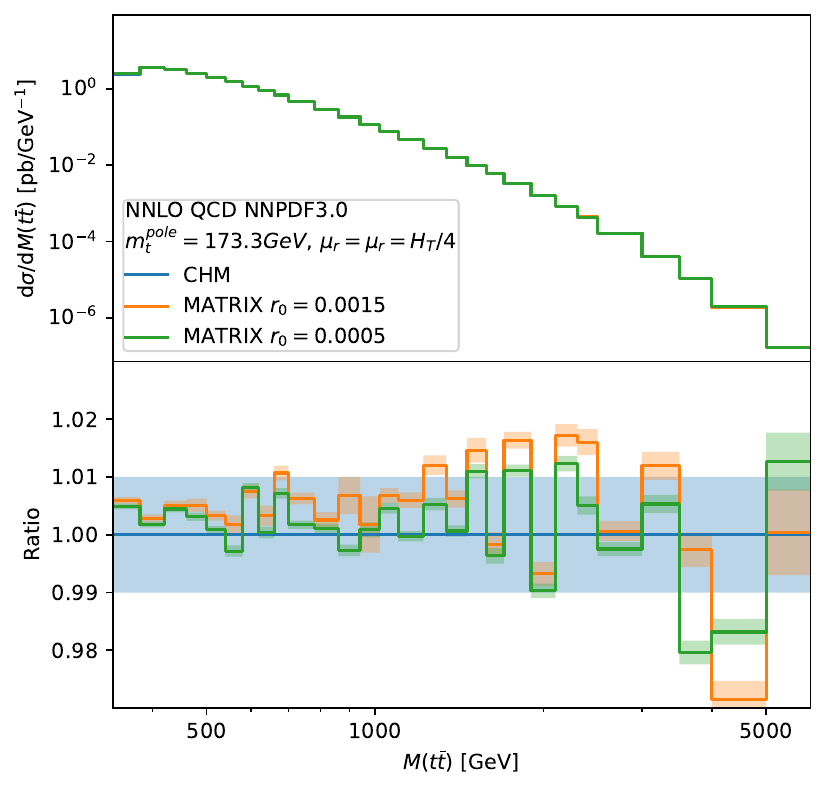}
    \includegraphics[width=0.49\textwidth]{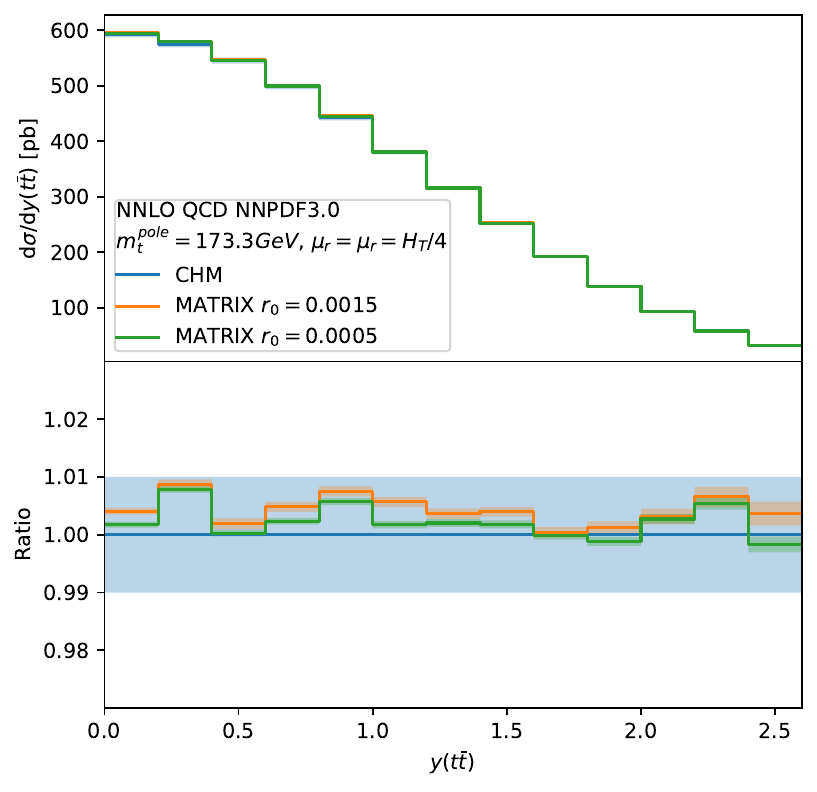}
    \caption{Comparison between the NNLO
      differential cross sections as a function of a) the average transverse momentum of the top quark and antiquark (upper left), b) the average rapidity of the top quark and antiquark (upper right),  c) the invariant mass of the top-antitop       quark pair (lower left) and d) the rapidity of the top-antitop 
      quark pair         (lower right), computed with \texttt{MATRIX} using the cuts $r_0=0.0015$ and $r_0=0.0005$, and the results from Ref.~\cite{Czakon:2016dgf}, 
      dubbed CHM in the figure,
      with their numerical uncertainties.}
    \label{fig:chm-rcut}
\end{figure}

\begin{figure}[htb]
    \centering
    \includegraphics[width=1.00\textwidth]{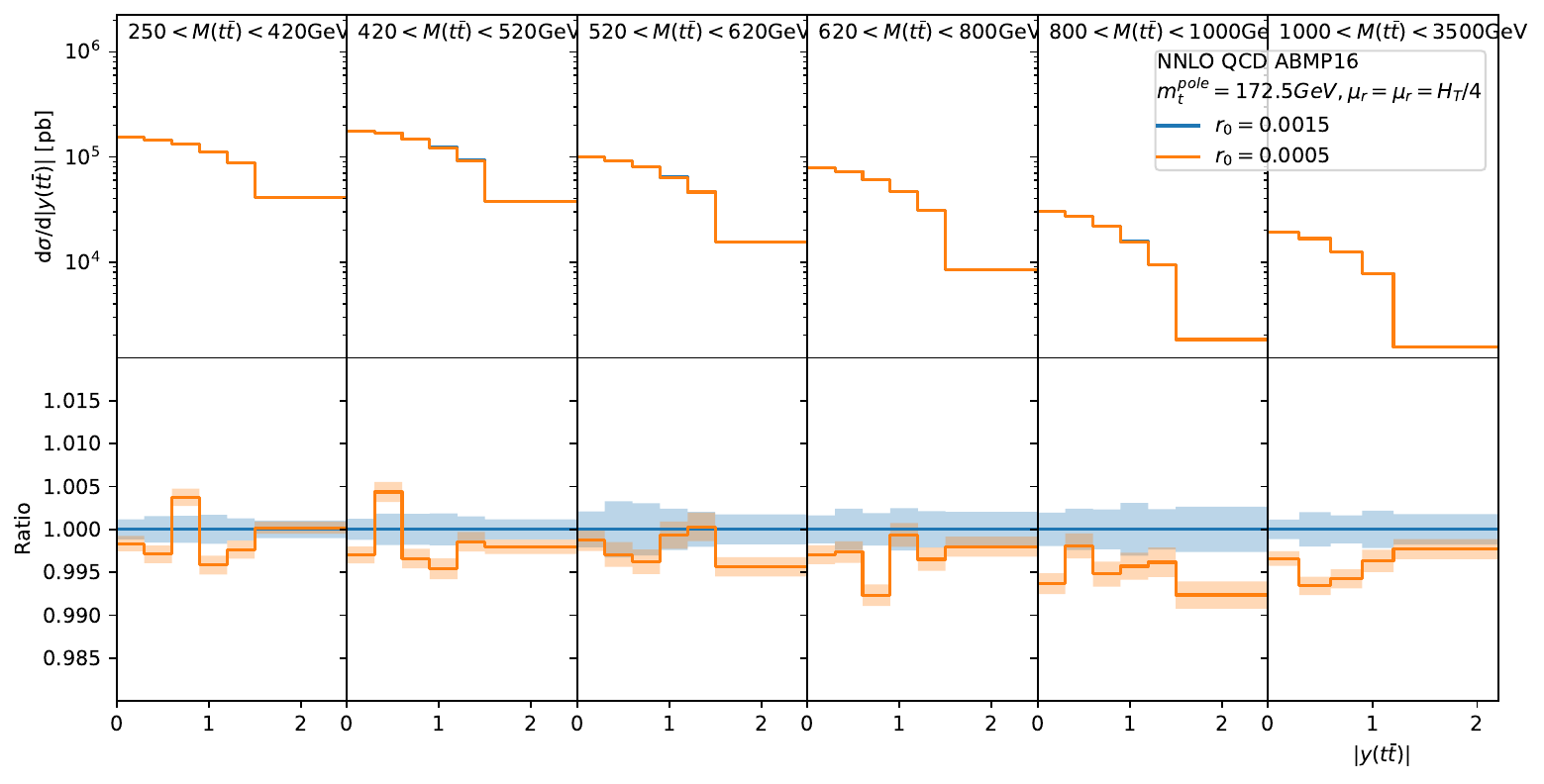}
    \caption{Comparison between the NNLO
      differential cross sections as a function of the rapidity of the top-antitop       quark pair in different ranges of the invariant mass of the top-antitop       quark pair, computed with \texttt{MATRIX} using the cuts $r_0=0.0015$ and $r_0=0.0005$.}
    \label{fig:rcut-dd}
\end{figure}

\smallskip
In Fig.~\ref{fig:chm-rcut} we compare the differential cross sections at NNLO we have computed with \texttt{MATRIX}, after the customization and optimization steps described above, with those from Ref.~\cite{Czakon:2016dgf}.
For the \texttt{MATRIX} framework, we compare the differential cross sections computed using the cuts $r_0=0.0015$ and $r_0=0.0005$~\footnote{As a baseline, in our work we use the default value $r_0=0.0015$ since it provides a faster convergence of the calculation. The statistics of the numerical calculations in the high-energy tails of the distributions are enhanced by setting \texttt{optimization\_modifier=1}~\cite{matrixmanual}.}.
For this calculation we use the NNPDF3.0 PDF set~\cite{NNPDF:2014otw} and a top-quark pole mass value $\mtpole=173.3$~GeV. When computing the cross sections as a function of the transverse momentum of either the top or the antitop quark, the factorization and renormalization scales are set to $\mu_r = \mu_f = \sqrt{m_t^2+p_{T,t}^2}/2$, i.e.\ to a half of the transverse mass of either the top or the antitop quark, while for all other observables the scales are set to $\mu_r = \mu_f = H_T/4$ where $H_T$ is the sum of the transverse masses of the top and the antitop quarks, consistent with Ref.~\cite{Czakon:2016dgf}.
While no numerical uncertainties are provided for the results from Ref.~\cite{Czakon:2016dgf}, we roughly estimate them to be $1\%$, based on corresponding comments in the text therein, as done also in Ref.~\cite{Catani:2019hip}, 
for the comparison of the NNLO predictions 
against those of Ref.~\cite{Czakon:2016dgf}. The results of the two computations agree within $\approx 1\%$. 
No trends are observed apart from a very small $p_T(t)$ slope ($< 1\%$) and the general normalization: the \texttt{MATRIX} $r_0 = 0.0015$ predictions are $\approx 0.5\%$ above those from Ref.~\cite{Czakon:2016dgf}. 
This could be related to the fact that, in the version of the code used in this study, the differential distributions are computed with finite cuts $r_0 = 0.0015$ and $r_0 = 0.0005$. 
For the total $\ttbar + X$ cross section \texttt{MATRIX} performs automatically an extrapolation to $r_0 = 0$, which results in a systematic correction of similar size, $-0.5\%$, thus suggesting power corrections in $q_T$ as a source for the observed difference~\footnote{We note that the latest public version of \texttt{MATRIX} allows the use of the bin-wise extrapolation. We did not consider this feature in our analysis since, as discussed in the main text, it would have no noticeable effect in our results, and its use would imply a significant increase in disk usage. In addition, in Ref.~\cite{Catani:2019hip} it was found that the distributions
extrapolated to $r_0 = 0$ are in good agreement with those obtained using a sufficiently small cut ($r_0 \lesssim 0.0015$).}.
In Fig.~\ref{fig:rcut-dd} we present the comparison of \texttt{MATRIX} predictions for the double-differential cross section as a function of \mtt and \ytt obtained with $r_0 = 0.0015$ and $r_0 = 0.0005$, using the binning scheme of the experimental measurement~\cite{top20001}. The results agree within $\approx 0.5\%$ for all distributions. 
For the phenomenological analysis this small difference does not create any issue, since an effect of the size of $0.5\%$ is very well covered by e.g. the scale variation uncertainties which we do take into account (see Section~\ref{sec:prefit}). 
Additionally, this effect partially cancels between numerator and denominator when considering normalized differential cross sections, that form the basis of our fits, as discussed in Sections~\ref{sec:compa} and \ref{sec:fit}.

\begin{figure}[htb]
    \centering
    \includegraphics[width=0.49\textwidth]{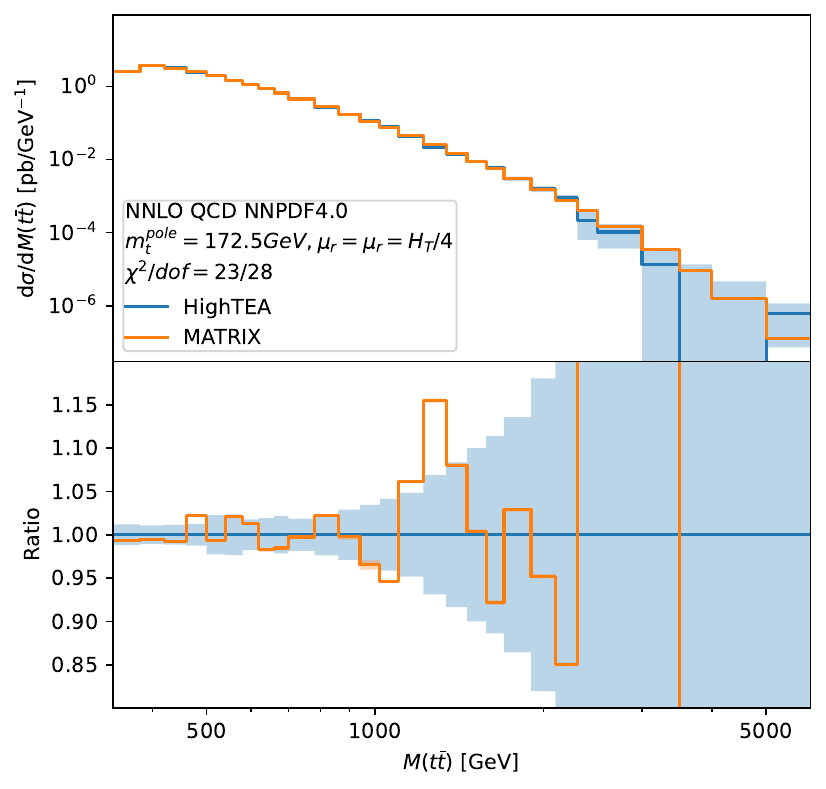}
    \caption{Comparison of the NNLO
      differential cross sections as a function of \mtt     computed with \texttt{MATRIX} with       the results obtained using the \texttt{HighTEA} project
      platform~\cite{Czakon:2023hls}. The error bars account for the numerical uncertainties accompanying the two computations.
\label{fig:hightea}
}
\end{figure}

Furthermore, in Fig.~\ref{fig:hightea} we compare the NNLO differential cross sections as a function of \mtt computed with \texttt{MATRIX} to the results obtained using the recently developed \texttt{HighTEA} project platform~\cite{Czakon:2023hls}. 
For this calculation we use the NNPDF4.0 PDF set~\cite{NNPDF:2021njg}, $\mtpole=~172.5$~GeV and $\mu_r = \mu_f = H_T/4$. Unfortunately, the \texttt{HighTEA} theoretical predictions are currently accompanied by quite large numerical uncertainties which vary from $1\%$ to $\sim 100\%$ at large \mtt values.
The two calculations agree within these uncertainties. We note that the current level of accuracy for differential $\ttbar+X$ production, that can be reached through the \texttt{HighTEA} project platform,  
is not sufficient for a phenomenological analysis of the present LHC data. We expect that this can indeed be improved by enlarging the event files stored there to include a higher number of events.  
Therefore, for the time being, in the following of this work we limit ourselves to the use of \texttt{MATRIX}. 

\subsection{Interface to \texttt{PineAPPL}}
\label{sec:pineappl}

A target precision of a few per mill accuracy requires the generation of various billions of $t\bar{t} + X$ NNLO events, which takes $\mathcal{O}(10^5)$ CPU hours. 
Repeating such a computation for many different PDF parametrizations and/or different scale choices is not feasible with present standard CPU computing resources available to the theory and experimental HEP communities. 
While the possibility to store the calculation results for a few different scale choices is already available within the \texttt{MATRIX} framework, in a single run there is neither the possibility to compute predictions using different PDF sets nor different error members within a single PDF set, at least for the time being.
A general solution to this problem is to use interpolation grids, such as \texttt{fastNLO}~\cite{Britzger:2012bs}, \texttt{APPLgrid}~\cite{Carli:2010rw} or \texttt{PineAPPL}~\cite{Carrazza:2020gss}.
In these grids, partonic matrix elements are stored in such a way that they can be convoluted later with any PDF + $\alpha_s(M_Z)$ set. We choose the \texttt{PineAPPL} library which is capable of generating grids and dealing with them in an accurate way to any fixed order in the strong and electroweak couplings, and which supports variations of $\mu_r$ and $\mu_f$. 
For each event computed by \texttt{MATRIX}, we fill the \texttt{PineAPPL} grid for the given perturbative QCD contribution and parton luminosity with the corresponding event weight divided by the product of the two proton PDFs involved in the event-weight computation. 
Cross-section terms which have different $\mu_r$ and/or $\mu_f$ dependence are stored in separate grids.
The \texttt{PineAPPL} grids are three-dimensional Lagrange-interpolation grids binned in variables which represent the longitudinal momentum fractions for the two incoming partons and $\mu_f$.
The values stored in each grid are independent of $\alpha_s$ and PDFs. Further details can be found in Ref.~\cite{Carrazza:2020gss}.

\begin{figure}[h!]
    \centering
    \includegraphics[width=1.0\textwidth]{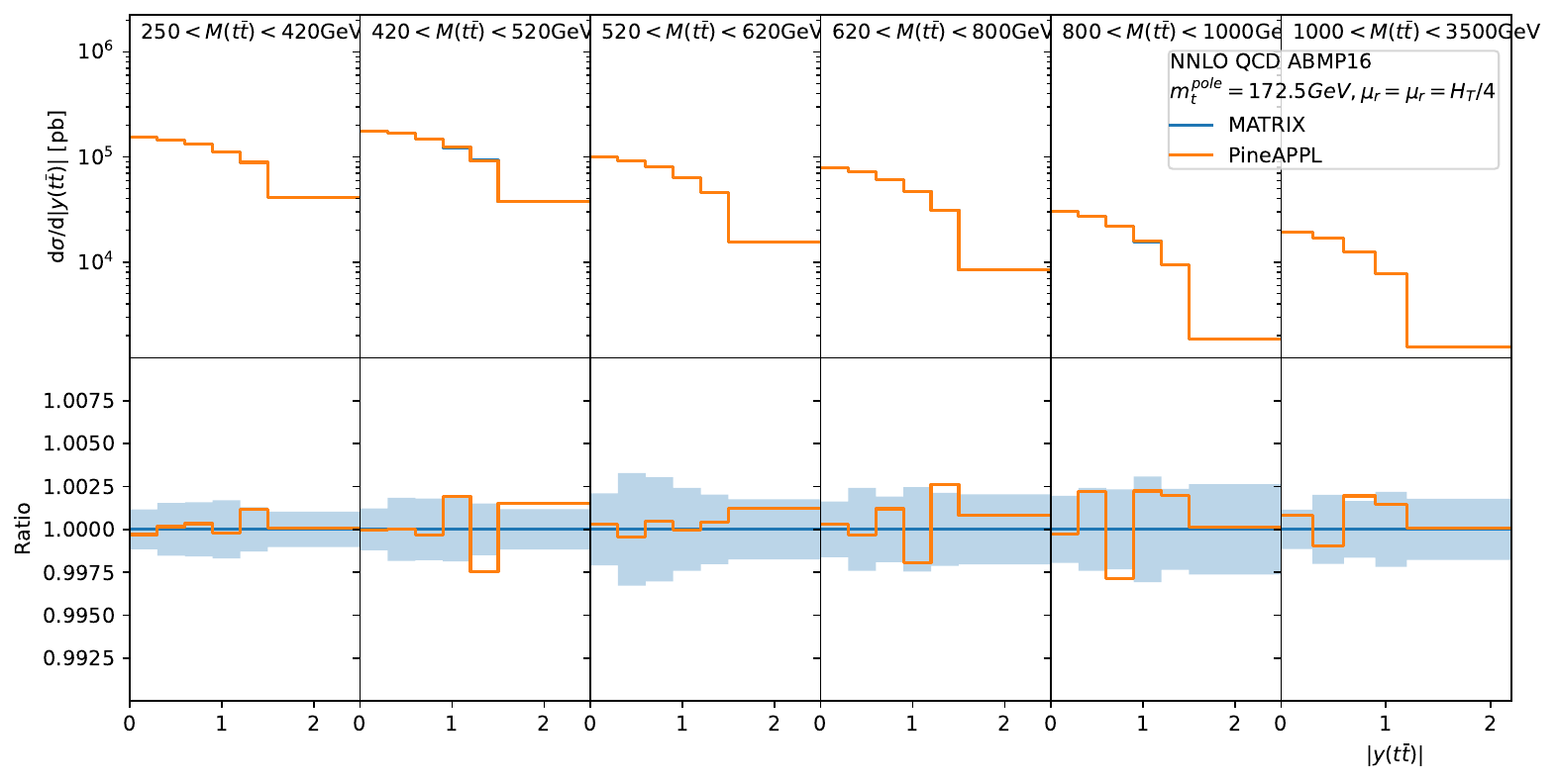}
    \includegraphics[width=1.0\textwidth]{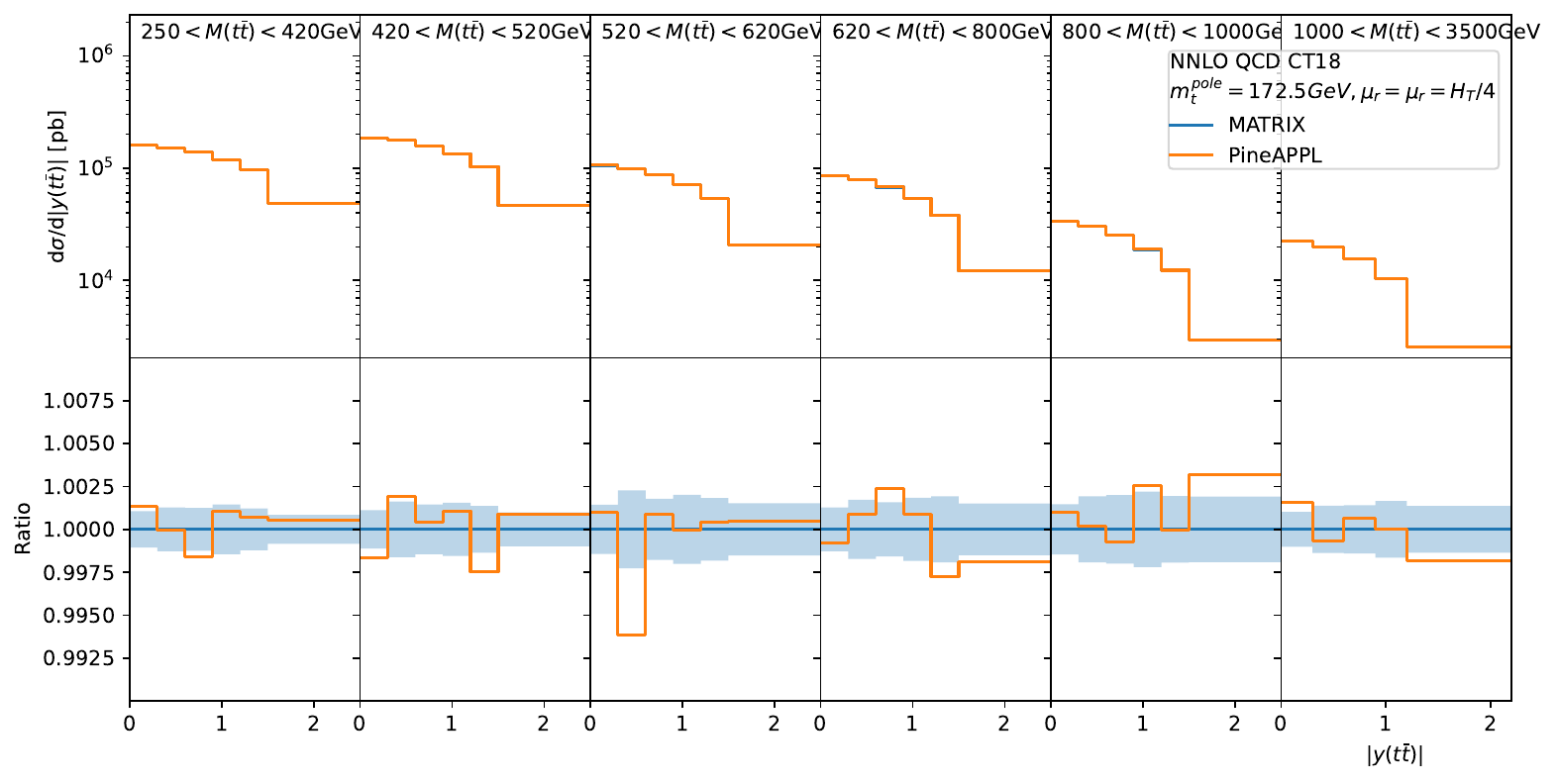}
    \caption{Comparison of the  NNLO
      differential cross sections with the ABMP16 (upper row) and the CT18 (lower row) PDFs      as a function of \ytt
      computed with \texttt{MATRIX}
      with their numerical uncertainties and the ones obtained from the
      \texttt{PineAPPL} grids. Each panel in a row
      refers
      to a different \mtt region, as in            
      the experimental measurement from Ref.~\cite{top20001}. 
    \label{fig:matunc}}
\end{figure}

To produce the grids, we have chosen $30$ bins for each internal variable of the grids. 
In order to validate our implementation of the interface to \texttt{PineAPPL}, in Fig.~\ref{fig:matunc} we compare the genuine theoretical predictions from \texttt{MATRIX} with those obtained using the \texttt{PineAPPL} interpolation grids (derived by employing only the ABMP16 NNLO PDF + $\alpha_s(M_Z)$ fit~\cite{Alekhin:2017kpj} as input of the \texttt{MATRIX} computation) and either the ABMP16 or the CT18 NNLO~\cite{Hou:2019efy} PDF + $\alpha_s(M_Z)$ sets. 
We do this comparison in the \mtt and \ytt bins of the experimental measurement from Ref.~\cite{top20001}, which, among all the experimental analyses producing the data which we use in this work, has the largest number of bins and covers the largest phase space. 
For the \texttt{MATRIX} calculation used to produce the grids, we require $0.2\permil$ accuracy for the total \ttbar~+~$X$ cross section.
This results in numerical uncertainties which are also shown in Fig.~\ref{fig:matunc}.
They vary from $0.1\%$ to $0.3\%$ for the bins of the experimental measurement of Ref.~\cite{top20001}. 
The typical difference between the original \texttt{MATRIX} predictions and those obtained using the \texttt{PineAPPL} grids turns out to be $\approx 1 \permil$ for the ABMP16 set and a few $\permil$ for the CT18 set for each event.
As expected, the \texttt{PineAPPL} interpolation uncertainty is very small if the same PDF set is used to produce the grids. The slight increase when using a different PDF set is related to the fact that the grids are organized in bins of finite size.
On the other hand, when accumulating a big number of events, for the ABMP16 set the differences remain of the order of $\permil$, whereas for the other set (CT18) the differences never exceed $1\%$ and are consistent with the \texttt{MATRIX} numerical uncertainties, since in this case two independent calculations are compared. 
The same level of agreement can be expected for any other reasonably smooth PDF set.
Thus, the usage of interpolation grids does not deteriorate the accuracy of our \texttt{MATRIX} calculation with the current numerical uncertainties.
Whenever necessary (e.g.\ for a finer binning scheme), the level of agreement can be improved further by producing new interpolation grids with an increased number of bins for the internal variables, however, at the price of an increased computer memory and disk space needed for the grids (currently it is more than one GByte and it varies depending on multiple factors, including among others not only the number of bins, but even the number of observables and parallel processes, considering that all runs are parallelized).

Based on these validation studies, we assign $1\%$ uncorrelated uncertainty in each bin of the predictions in order to cover any numerical inaccuracy and possible small systematic effects which might be present either in the theoretical calculations (see Figs.~\ref{fig:chm-rcut} and \ref{fig:rcut-dd}) or in the usage of the interpolation grids (see Fig.~\ref{fig:matunc}). This uncertainty is always smaller than the experimental uncertainties of the existing single- and double-differential measurements of $\ttbar+X$ production at the LHC which typically amount to a few percent (although the 1\% additional uncertainty is sizeable, e.g., in some bins of the data from Ref.~\cite{top20001}).

\section{Pre-fit comparison of theory predictions and LHC experimental data}
\label{sec:compa}

\subsection{Experimental data}
\label{sec:data}

For our analysis, we use measurements of the absolute total and normalized differential inclusive \ttbar~+~$X$ cross sections\footnote{In this work we use the word ``inclusive'' as synonymous of ``without analysis cuts'', i.e., extrapolated to the full phase space, considering that both the ATLAS and the CMS experimental collaborations already performed this extrapolation when providing cross sections for $t\bar{t}+X$ production. The extrapolation uncertainties are included in the experimental error bars.} as a function of various top-quark related observables $O_i$, $(d\sigma/dO_i)/\sigma$.
We collect all available and up-to-date ATLAS and CMS measurements of total \ttbar + $X$ cross sections~\cite{tca78,ta5,tc5,top20001,tc13ll,ta13lj,ta13ll,ta136,tc136}. These are ten data points which appear on the summary plot of the total \ttbar + $X$ cross sections by the LHC Working Group on Top Quark Physics as of June 2023~\cite{lhctopwg}. They correspond to measurements at $\sqrt{s} = 5.02,\,7,\,8,\,13$ and $13.6$~TeV, as summarized in Table~\ref{tab:data_total}.

\begin{table}[htb]
\centering
\begin{tabular}{llllll}
    experiment & decay channel & dataset & luminosity & $\sqrt{s}$ & ref.\\ \hline
    ATLAS \& CMS & combined & 2011 & $5$ fb$^{-1}$ & 7 TeV & \cite{tca78} \\
    ATLAS \& CMS & combined & 2012 & $20$ fb$^{-1}$ & 8 TeV & \cite{tca78} \\
    ATLAS & dileptonic, semileptonic & 2011 & $257$ pb$^{-1}$ & 5.02 TeV & \cite{ta5} \\
    CMS & dileptonic & 2011 & $302$ pb$^{-1}$ & 5.02 TeV & \cite{tc5} \\
    ATLAS & dileptonic & 2015-2018 & $140$ fb$^{-1}$ & 13 TeV & \cite{ta13ll} \\
    ATLAS & semileptonic & 2015-2018 & $139$ fb$^{-1}$ & 13 TeV & \cite{ta13lj} \\
    CMS & dileptonic & 2016 & $35.9$ fb$^{-1}$ & 13 TeV & \cite{tc13ll} \\
    CMS & semileptonic & 2016-2018 & $137$ fb$^{-1}$ & 13 TeV & \cite{top20001} \\
    ATLAS & dileptonic & 2022 & $11.3$ fb$^{-1}$ & 13.6 TeV & \cite{ta136} \\
    CMS & dileptonic, semileptonic & 2022 & $1.21$ fb$^{-1}$ & 13.6 TeV & \cite{tc136} \\
\end{tabular}
\caption{The measurements of total inclusive \ttbar +~$X$ cross sections included in our analysis.}
\label{tab:data_total}
\end{table}

On the other hand, for differential measurements, we choose cross sections as a function of the invariant mass of the \ttbar pair, or (if available) double-differential cross sections as a function of the invariant mass \mtt and rapidity \ytt of the \ttbar pair.
In particular, the $\ttbar + X$ cross sections as a function of \mtt are very useful to constrain \mtpole, while double-differential $\ttbar+X$ cross sections as a function of \mtt and \ytt impose constraints on the PDFs owing to the improved resolution of parton momentum fractions $x_1$ and $x_2$. Indeed, at LO the invariant mass \mtt and rapidity \ytt of the \ttbar pair are directly related to $x_1$ and $x_2$ as $x_{1,2} = (\mtt/\sqrt{s})\exp{[\pm y(\ttbar)]}$. 
Therefore, measurements of double-differential cross sections as a function of \mtt and \ytt are most sensitive to the PDFs and provide strong constraints both on \mtpole and the PDFs. 
Furthermore, we choose only those differential datasets which satisfy all of the following criteria:
\begin{itemize}
    \item measured cross sections should be defined at parton level in the full phase space, i.e.\ without any restrictions on the decay products of the top and antitop quarks, in order to be consistent with the parton-level NNLO calculations,
    \item normalized cross sections must be available, in order to avoid a complicated treatment of the common normalization with the datasets for the total $\ttbar + X$ cross sections,
    \item bin-by-bin correlations must be available.
\end{itemize}
Under these criteria, we use the nine datasets listed in Table~\ref{tab:data_diff}. In total, they contain $112$ data points, taking into account that one data point from each measurement should be discarded during the comparison of normalized data and theory, as explained later in Section~\ref{sec:chisq}. 

In our work, we extract the top quark pole mass, \mtpole, by comparing the measured $\ttbar+X$ cross sections to NNLO theoretical predictions which depend on \mtpole.
In order to measure the $\ttbar+X$ cross sections, experimental collaborations use an unfolding procedure. 
The aim of the unfolding procedure is to obtain the $\ttbar+X$ cross sections at parton level using as input various kinematic distributions at detector level. 
While the unfolding procedure uses MC simulations, by construction it 
{is supposed to provide}
$\ttbar+X$ cross sections at parton level which do not explicitly depend on the details of the MC simulations. 
This is {verified} by various checks, such as that of control distributions when one compares event distributions in the data and in the MC simulations, as well as closure tests, when, using toy distributions, one validates
that the unfolding procedure restores the truth reasonably well independently of the MC simulations used.
Any remaining dependence on the MC simulations is supposed to be covered by systematic uncertainties which are estimated e.g. by varying theoretical parameters in the MC simulations including, but not limiting to, the MC mass, {and} by adopting alternative models/tunes. 
With such a method, the measured parton-level cross sections are meant to be directly compared to fixed-order predictions, which is often done also by the experimental collaborations in their original publications (see e.g. Refs.~\cite{top14013,top18004,top20001,a190807305}).
{
This is different from other determinations of the top-quark mass at the LHC that rely upon the reconstruction of the top-quark from its decay products, and are usually dubbed ``direct measurements'' (see e.g.\ the recent review of the top-quark mass measurements in CMS~\cite{CMS:2024irj}).
The results obtained in direct measurements of the MC top-quark mass are affected by ambiguities  originating from theoretical uncertainties and limitations of the current MC simulations.
Nevertheless, in the case of \ttbar pairs produced in hadron-hadron collisions, where the underlying hard interactions unavoidably involve partons in non-singlet color configurations, the conceptual issues that affect the direct measurements are eventually emerging for all top-quark mass measurement methods, once a precision of 0.5~GeV or better is reached~\cite{Hoang:2020iah}. 
In view of this, we regard the present determination of the top-quark mass as the extraction of \mtpole from the differential distributions at the parton-level obtained by the experimental collaborations from those at the detector-level using MC event generators, for identifying in a more precise way the extraction of top-quark mass values that we performed. This extended formulation accounts for the fact that there might be a residual implicit dependence of our extracted top-quark mass value (merely dubbed as \mtpole throughout the text), on the MC top-quark mass, $m_t^{\text{MC}}$, although the experimental collaborations did their best to minimize the dependence on $m_t^{\text{MC}}$ of the unfolded data at the basis of our fits. Further investigations concerning this dependence go beyond the scope of the present work and might require access to the detector-level experimental data, not publicly available, in order to study how to further improve the unfolding procedure.}

{Furthermore}, one should distinguish a situation when some unfolded data distribution might strongly depend on a parameter of the MC simulations (such as $m_t^{\text{MC}}$), as it happens to the differential cross section as a function of the invariant mass of the \ttbar pair in the dilepton channel when using the so called full kinematic reconstruction (due to a peculiar aspect of the kinematic reconstruction, namely because of the need to use the top-quark mass constraint at the reconstruction level, $m_t^{\text{kin}}$, in order to deal with the unknown momenta of the neutrinos which escape the detector).
This dependence is covered by changing the value of $m_t^{\text{kin}}$ and of $m_t^{\text{MC}}$ (typically varied independently within $m_t^{\text{kin}}=172.5 \pm 1$ GeV, $m_t^{\text{MC}}=172.5 \pm 1$ GeV, see e.g. Refs.~\cite{top14013,top18004}).
These variations are included in the systematic uncertainties accompanying the measured cross sections.
If, however, the real value of these two parameters is outside the range of considered variations, it would introduce a bias.
This $m_t^{\text{kin}}$ constraint is not used in the loose kinematic reconstruction developed in the CMS measurement from Ref.~\cite{top18004}. As a consequence, for this measurement the dependence of the measured \ttbar~+~$X$ cross sections on $m_t^{\text{MC}}$ was demonstrated to be negligible (see Fig.~C.1 in Ref.~\cite{top18004}), 
cf. also Section~\ref{sec:prefit}.
%

\begin{table}[htb]
\setlength{\tabcolsep}{5pt}
\centering
\begin{tabular}{llllllll}
    Experiment & decay channel & dataset & luminosity & $\sqrt{s}$ & observable(s) & $n$ & ref.\\ \hline
    CMS & semileptonic & 2016--2018 & $137$ fb$^{-1}$ & 13 TeV & \mtt, \ytt & 34 & \cite{top20001} \\    CMS & dileptonic & 2016 & $35.9$ fb$^{-1}$ & 13 TeV & \mtt, \ytt & 15 & \cite{top18004} \\
    ATLAS & semileptonic & 2015--2016 & $36$ fb$^{-1}$ & 13 TeV & \mtt, \ytt & 19 & \cite{a190807305} \\
    ATLAS & all-hadronic & 2015--2016 & $36.1$ fb$^{-1}$ & 13 TeV & \mtt, \ytt & 10 & \cite{a200609274} \\
    CMS & dileptonic & 2012 & $19.7$ fb$^{-1}$ & 8 TeV & \mtt, \ytt & 15 & \cite{top14013} \\
    ATLAS & semileptonic & 2012 & $20.3$ fb$^{-1}$ & 8 TeV & \mtt & 6 & \cite{a151104716} \\
    ATLAS & dileptonic & 2012 & $20.2$ fb$^{-1}$ & 8 TeV & \mtt & 5 & \cite{a160707281} \\
    ATLAS & dileptonic & 2011 & $4.6$ fb$^{-1}$ & 7 TeV & \mtt & 4 & \cite{a160707281} \\
    ATLAS & semileptonic & 2011 & $4.6$ fb$^{-1}$ & 7 TeV & \mtt & 4 & \cite{a14070371} \\
\end{tabular}
\caption{The measurements of differential inclusive \ttbar + $X$ cross sections included in our analysis. The number of data points ($n$) reported does not account for the last data bin, considering that the latter is discarded for all measurements in the data-to-theory comparison process.}
\label{tab:data_diff}
\end{table}

Since we had to limit our choice to the measurements done in the full phase space, we do not use any of the LHCb measurements of \ttbar +~$X$ production~\cite{LHCb:2015nta,LHCb:2016hnm,LHCb:2018usb}, which provide \ttbar~+~$X$ cross sections with cuts on the \ttbar decay products. In the future, any measurements of \ttbar production at LHCb reported after unfolding to the top-quark level without explicit cuts on the decay products would be useful.

\subsection{\chisq definition}
\label{sec:chisq}

The level of agreement between data and theory can be quantified using the \chisq estimator. A $\chi^2$ value is calculated by taking into account statistical and systematic experimental uncertainties as well as theoretical uncertainties:
\begin{equation}
\label{eq:chi2nm1}
\chisq = \mathbf{R}^{T}_{N-1} \mathbf{Cov}^{-1}_{N-1} \mathbf{R}_{N-1}.
\end{equation}
Here $\mathbf{R}_{N-1}$ is the column vector of the residuals calculated as the difference of the measured cross sections and theoretical predictions obtained by discarding one of the $N$ bins (see below), and $\mathbf{Cov}_{N-1}$ is the $(N-1)\times(N-1)$ submatrix obtained from the full covariance matrix by discarding the corresponding row and column.
The matrix $\mathbf{Cov}_{N-1}$ obtained in this way is invertible, while the original covariance matrix $\mathbf{Cov}$ is singular because for normalised cross sections one degree of freedom is lost.
The covariance matrix $\mathbf{Cov}$ is calculated as:
\begin{equation}
\label{eq:covmat}
\mathbf{Cov} = \mathbf{Cov}^\text{stat} + \mathbf{Cov}^\text{syst}
+ \mathbf{Cov}^\text{th} + \mathbf{Cov}^\text{PDF},
\end{equation}
where $\mathbf{Cov}^\text{stat}$ and $\mathbf{Cov}^\text{syst}$ are the covariance matrices corresponding to the statistical and systematic uncertainties reported in the experimental papers, respectively, $\mathbf{Cov}^\text{th}$ consists of numerical uncertainties of the theoretical predictions, and $\mathbf{Cov}^\text{PDF}$ is the covariance matrix which comprises the PDF uncertainties.
For some of the datasets~\cite{top18004,top14013}, a detailed breakdown of systematic uncertainties into individual sources is reported in the corresponding paper,
instead of the covariance matrix.
In such a case, the systematic covariance matrix $\mathbf{Cov}^\text{syst}$ is calculated as
\begin{equation}
    \mathbf{Cov}^\text{syst}_{ij} = \sum_{k,l} \frac{1}{N_k} C_{j,k,l}C_{i,k,l},\\ \quad 1 \le i,j \le N,
    \label{eq:chi2}
\end{equation}
where $C_{i,k,l}$ stands for the systematic uncertainty from variation $l$ of source $k$ in the $i$th bin, and $N_k$ is the number of variations for source $k$ (most of the systematic uncertainty sources consist of convenient positive and negative variations, i.e.\ $N_k = 2$), and the sums run over all sources of the systematic uncertainties and all corresponding variations. 
The covariance matrix $\mathbf{Cov}^\text{th}$ is a diagonal matrix which includes a $1\%$ uncorrelated uncertainty on the predictions, which we assign to cover any numerical inaccuracy and possible small systematic effects in the theoretical computations (see Section~\ref{sec:matrixtheory}).
In the same way, $\mathbf{Cov}^\text{PDF}$ is calculated from the variations of theoretical predictions obtained using individual PDF error members.
All uncertainties are treated as additive, i.e.\ the relative uncertainties are used to scale the corresponding measured value in the construction of $\mathbf{Cov}^\text{stat}$, $\mathbf{Cov}^\text{syst}$, $\mathbf{Cov}^\text{th}$ and $\mathbf{Cov}^\text{PDF}$.
This treatment is consistent with the cross-section normalization procedure and makes the \chisq independent of which of the $N$ bins is excluded.

No correlation has to be assumed between individual datasets, because no such information is provided by the experiments~\footnote{The exception is the recent combined CMS and ATLAS result for the total \ttbar$~+~X$
cross section at $\sqrt{s}=7$ and $8$~TeV~\cite{tca78}, 
for which such a correlation is reported and included in our analysis.}.
We deliberately opted to use normalized differential cross sections in order to minimize the impact of the lack of this information, since many correlated experimental systematic uncertainties cancel out for normalized cross sections.

\subsection{Comparison of the NNLO theoretical predictions with the experimental data}
\label{sec:prefit}

For our comparison of data and NNLO theoretical predictions, we use four state-of-the-art NNLO proton PDF sets 
as input of the theory computations:
ABMP16~\cite{Alekhin:2017kpj}, CT18~\cite{Hou:2019efy}~\footnote{In this work the PDF uncertainties of the CT18 set, evaluated at 90\% confidence level, are rescaled to 68\% confidence level for consistency with other PDF sets.}, that we already used for the validation/comparison of purely theoretical predictions in Section~\ref{sec:matrixtheory}, MSHT20~\cite{Bailey:2020ooq} and NNPDF4.0~\cite{NNPDF:2021njg}~\footnote{For the NNPDF4.0 set, we use its variant with eigenvector uncertainties in order to be able to calculate the $\mathbf{Cov}^\text{PDF}$ matrix in analogy to the other PDF sets.}. For each PDF set, we take the associated $\alpha_s(M_Z)$ value and $\alpha_s$ evolution via \texttt{LHAPDF}~\cite{Buckley:2014ana}. 
Namely, in the extraction of ABMP16 PDFs the value $\alpha_s = 0.1147 \pm 0.0008$ was obtained in the fit, while for the extraction of the CT18, MSHT20, and NNPDF4.0 PDF sets a fixed value of $\alpha_s = 0.118$ was used.
We remind here that the ABMP16 fit has incorporated data on total $t\bar{t} + X$ production cross section at the LHC Run~1, while the CT18, MSHT20 and NNPDF4.0 ones have also incorporated some single- or double-differential distributions for this process. 
For one of the PDF + $\alpha_s(M_Z)$ sets (ABMP), we consider different top-quark pole mass values ($\mtpole = 170, 172.5, 175$~GeV), as well as 7-point scale variation around a central renormalization and factorization scale $(\mu_R, \mu_F) = (\xi_R, \xi_F) \in \{(1, 1), (0.5., 0.5), (0.5,1), (1,0.5), (1, 2), (2, 1), (2, 2)\}(\mu_R^0, \mu_F^0)$, with the nominal scales $\mu_R^0=\mu_F^0= H_T/4$, where $H_T$ is the sum of the transverse masses of the top and the antitop quarks~\cite{Czakon:2016dgf,Catani:2019hip}.

\begin{figure}[htb]
    \centering
    \includegraphics[width=0.95\textwidth,clip,trim=0 10 0 6.5]{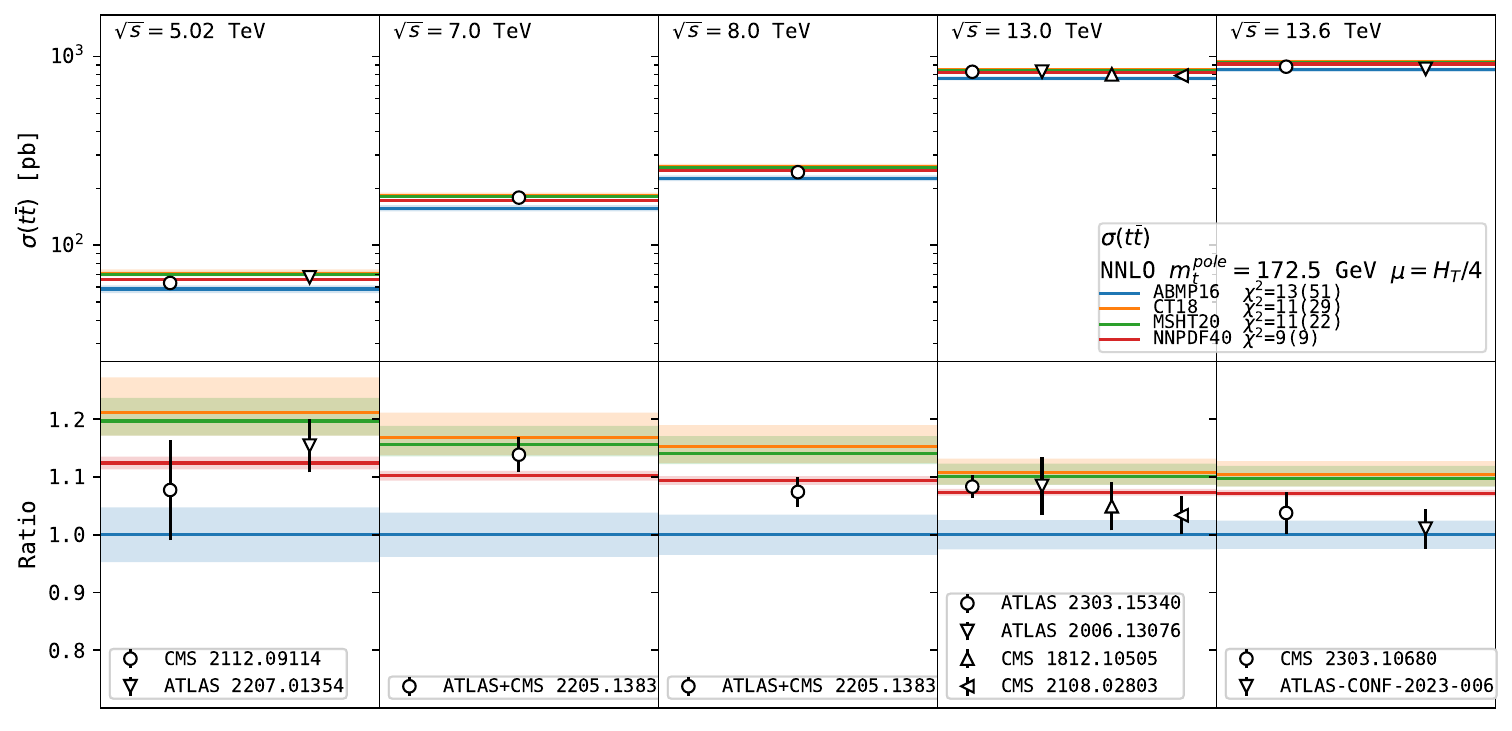}
    \includegraphics[width=0.95\textwidth,clip,trim=0 10 0 6.5]{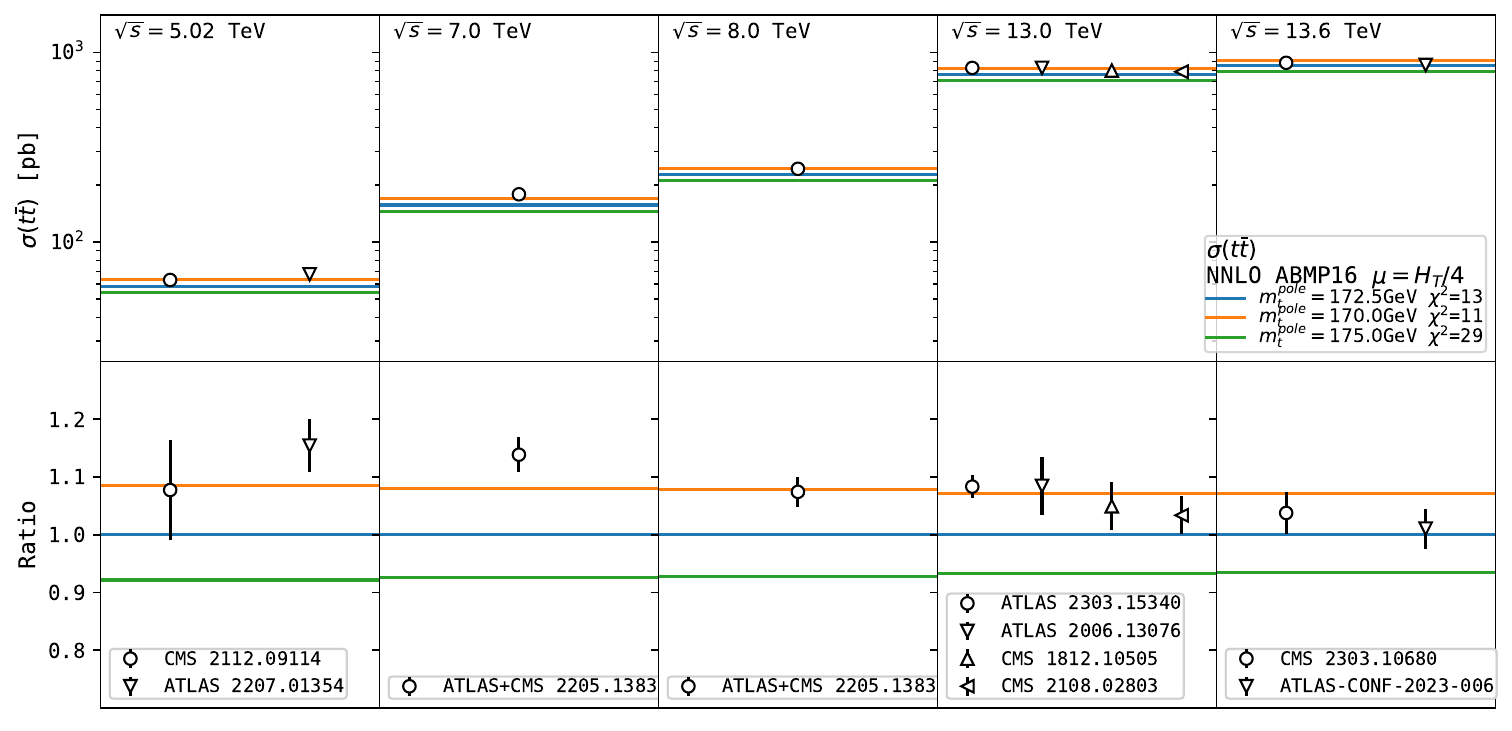}
    \includegraphics[width=0.95\textwidth,clip,trim=0 10 0 6.5]{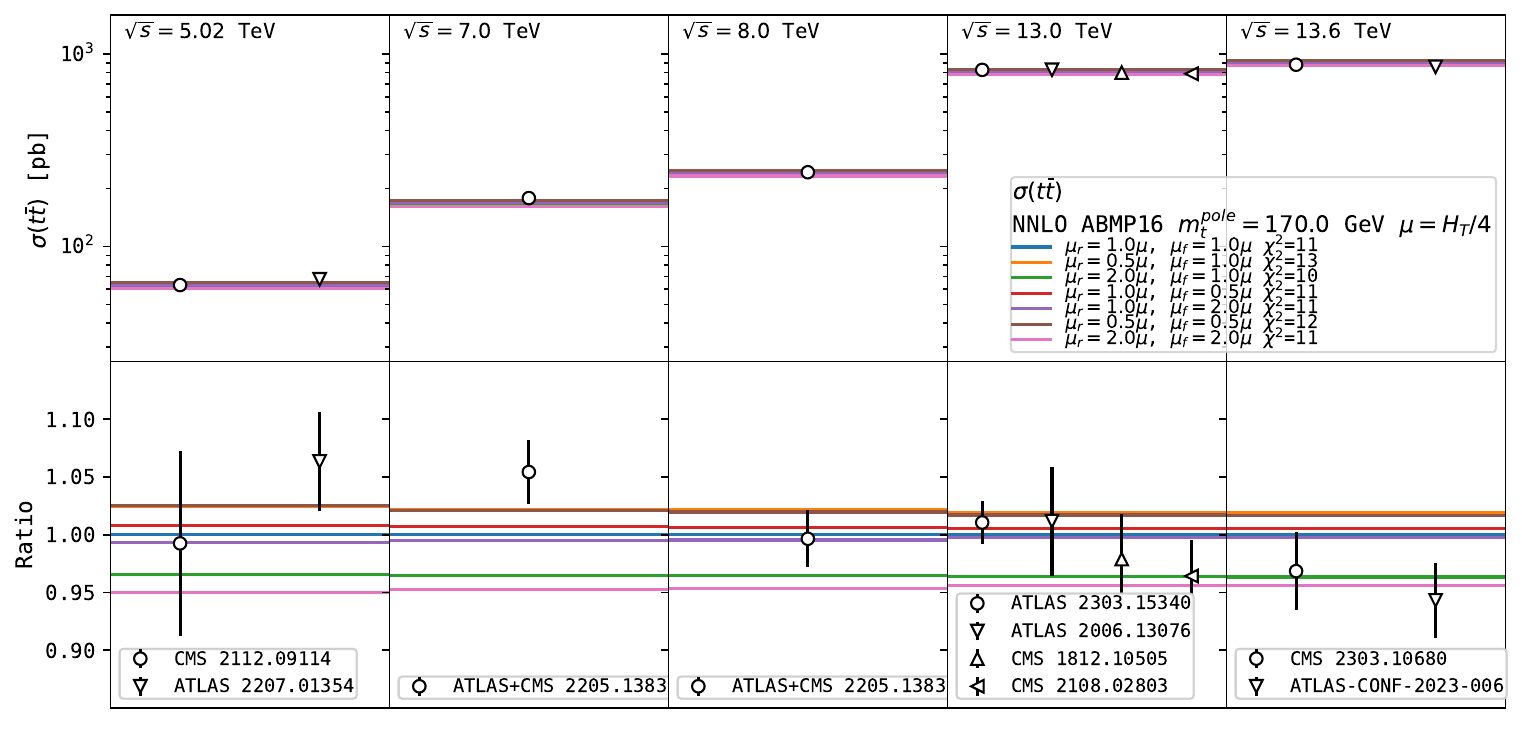}
    \caption{Comparison of the experimental data on the total \ttbar~+~$X$ cross sections at different $\sqrt{s}$  
from Refs.~\cite{tca78,ta5,tc5,top20001,tc13ll,ta13lj,ta13ll,ta136,tc136} to the NNLO predictions obtained using different PDF sets (upper), and, for the ABMP16 central PDF member, different \mtpole values (middle) and different scales (lower).}
    \label{fig:total}
\end{figure}

We start in Fig.~\ref{fig:total} by comparing the absolute total \ttbar~+~$X$ cross-sections at $\sqrt{s}$~=~5.02, 7, 8, 13 and $13.6$ TeV from Refs.~\cite{tca78,ta5,tc5,top20001,tc13ll,ta13lj,ta13ll,ta136,tc136} with the theoretical predictions. 
The first row of plots shows predictions obtained with different PDF~+~$\alpha_s(M_Z)$ sets, at fixed $\mtpole=~172.5$~GeV, the second row shows predictions for different \mtpole mass values, whereas the third row shows predictions with ABMP16 for different ($\xi_R, \xi_F$) combinations (see previous paragraph). 
For each comparison of data points and the corresponding theoretical predictions, a \chisq value is reported. In addition, for each PDF~+~$\alpha_s(M_Z)$ set, an additional \chisq value is provided (indicated in parentheses in the panels),  that omits the PDF uncertainties. 
Such \chisq values characterize the level of agreement between the data points and theoretical predictions obtained using the central PDF set only, while the difference between these \chisq values and the ones calculated with the PDF uncertainties provides an indication about the extent a given data sample could potentially constrain the PDF uncertainties of a particular PDF set.
From the first row of plots one can see that, within the PDF uncertainties, all considered PDF sets describe the data well.
The smallest PDF uncertainties occur for the NNPDF4.0 PDF set, followed by ABMP16, MSHT20 and CT18 PDF sets.
The sensitivity of the theoretical predictions to the PDF set decreases with increasing $\sqrt{s}$, since lower $\sqrt{s}$ probe larger values of $x$, and the large-$x$ region is characterized by a bigger PDF uncertainty, especially for the gluon PDF.
From the second row of plots, one can conclude that for the case of ABMP16 the value of \mtpole which is preferred by the data is between $170$ and $172.5$ GeV, while the larger value $\mtpole = 175$ GeV is clearly disfavoured.
Note that for the comparison with theoretical predictions which use different values of \mtpole or varied scales (i.e.\ for the second and third rows of plots) we do not show the PDF uncertainties on the plots for clarity, but we include them when calculating the \chisq values displayed on these plots.
As expected from kinematic considerations, the predicted \ttbar~+~$X$ cross sections decreases with increasing \mtpole.
Furthermore, it is noticeable that the sensitivity of the predicted cross section to \mtpole decreases slightly with increasing $\sqrt{s}$, since the mass then plays a smaller role in the kinematic region of the process.
In the third row of plots the behaviour of the predicted cross sections under different ($\xi_R$, $\xi_F$) combinations is shown, according to the 7-point scale variation around the central dynamical scale $\mu_0= H_T/4$. 
One can see that scale uncertainties are asymmetric, amount roughly to ${}^{+3}_{-5}\%$ and slightly decrease with increasing $\sqrt{s}$. 
Thus, the NNLO scale variation uncertainties for the total $\ttbar + X$ cross sections are larger than the experimental uncertainties of the most precise measurements of this process (e.g.\ in Ref.~\cite{ta13ll} the measured total \ttbar~$+ X$ cross section is reported with a $1.9\%$ total uncertainty).
No uncertainty associated with the scale dependence of the cross sections is included in the \chisq calculation because the scale variation uncertainty does not follow a Gaussian distribution, while for the extraction of \mtpole the scale uncertainties are propagated directly (see Section~\ref{sec:fit}).

\begin{figure}[htb]
    \centering
    \includegraphics[width=0.945\textwidth,clip,trim=0 41 0 6.5]{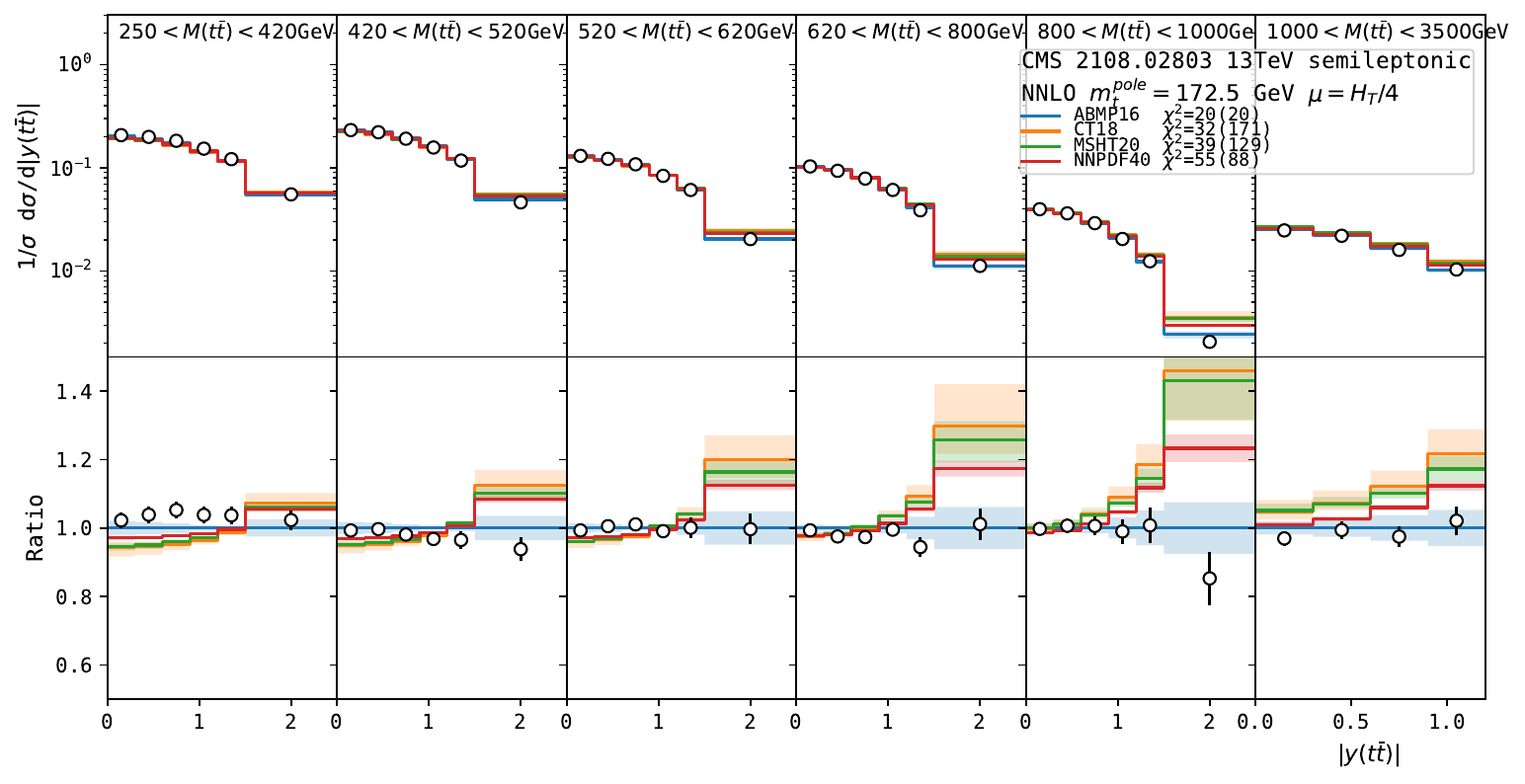}
    \includegraphics[width=0.945\textwidth,clip,trim=0 41 0 6.5]{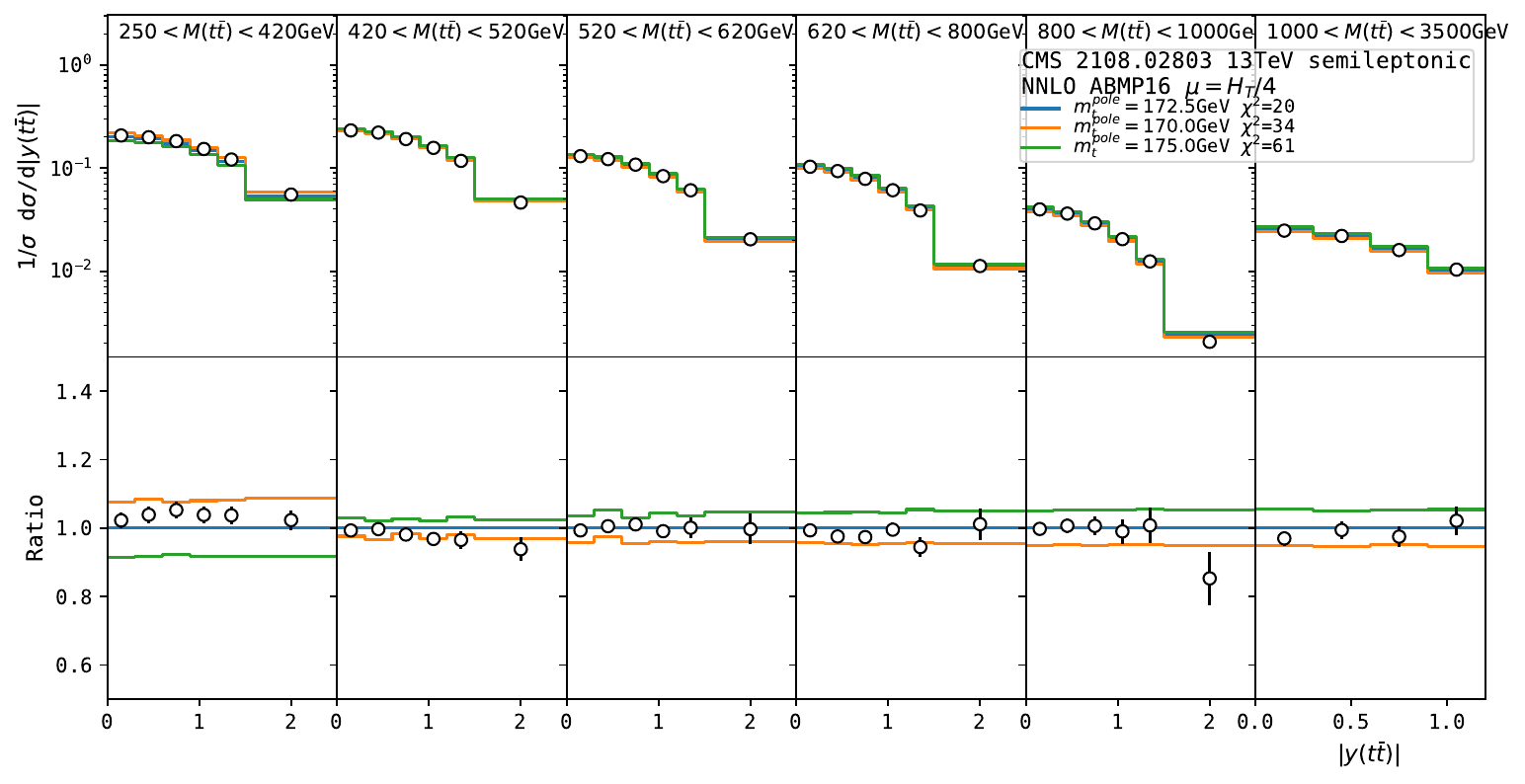}
    \includegraphics[width=0.945\textwidth,clip,trim=0 8 0 6.5]{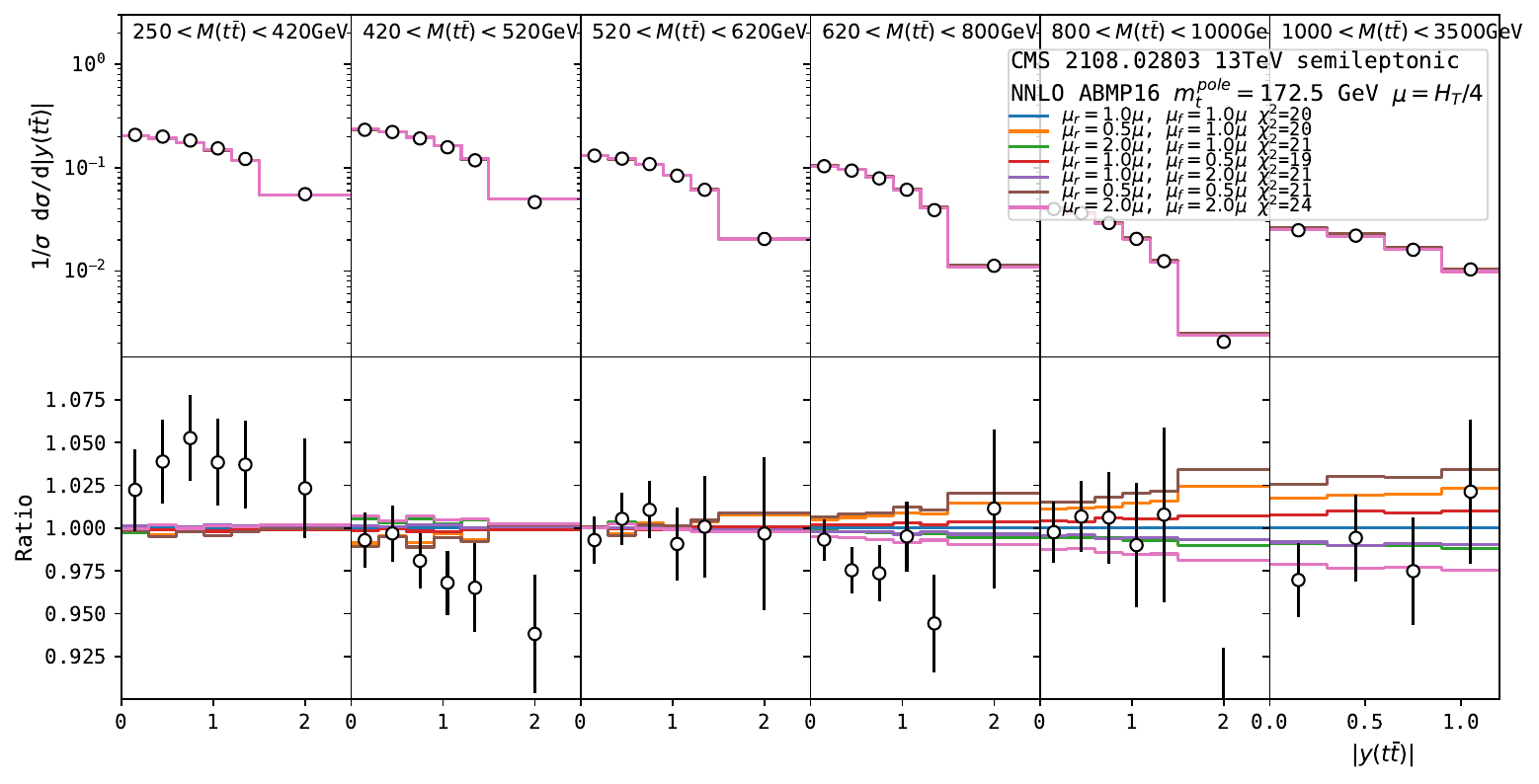}
    \caption{Comparison of the experimental data
from Ref.~\cite{top20001}
to the NNLO predictions obtained using different PDF sets (upper), and, for the ABMP16 central PDF member, different \mtpole values (middle) and different scales (lower).}
    \label{fig:top20001}
\end{figure}

After the discussion of the absolute total inclusive cross sections, we present results for the comparison of NNLO QCD theory predictions with normalized differential experimental data on inclusive $t\bar{t} + X$ hadroproduction.
The com\-pa\-ri\-sons are shown in Figs.~\ref{fig:top20001}--\ref{fig:a160707281_8tev}, which refer to the single- or double-differential experimental data of Refs.~\cite{top20001, top18004, a190807305, a200609274} obtained during Run~2, with the $t\bar{t}$-quark pair decaying in all possible channels (dileptonic, semileptonic, all-hadronic),  and of Refs.~\cite{top14013,a151104716,a160707281,a14070371} obtained during Run~1~\footnote{For all plots of the normalized differential cross sections, we do not show the last bin which is excluded from the \chisq calculation as explained in Section~\ref{sec:chisq}.}.

In Fig.~\ref{fig:top20001} the absolute value of the rapidity distribution of the $t\bar{t}$-quark pair, \ytt, is plotted in various $t\bar{t}$ invariant mass \mtt bins, corresponding to different panels, and compared to the experimental data of Ref.~\cite{top20001}, a CMS analysis with $t\bar{t}$-quark pairs decaying in the semileptonic channel. Considering the phase space of the measurement, the number of measured data points and their experimental uncertainties, this is presently the most precise LHC dataset for double-differential $\ttbar+X$ production cross section as a function of \mtt and \ytt, employing unfolding from the final-state particle to the final-state parton level. 
As in Fig.~\ref{fig:total}, the first row of plots shows predictions obtained with different PDF~+~$\alpha_s(M_Z)$ sets, at fixed $\mtpole = 172.5$~GeV, the second row shows predictions for different \mtpole mass values, whereas the third row shows predictions for different ($\xi_R, \xi_F$) combinations. 
As for the absolute total $\ttbar + X$ cross sections, each comparison of data to the corresponding theoretical predictions is characterized by a \chisq value. 
The first row clearly demonstrates that, in examining normalized cross sections,
the best agreement between theoretical predictions and experimental data for the shape
of the distributions is achieved when using the ABMP16 PDFs. 
Predictions with the CT18, MSHT20 and NNPDF4.0 show a similar trend among each other, but the shapes are systematically different from those of the experimental distribution at large \ytt, overestimating it. This is particularly evident in the large \mtt bins. 
As in the case of the total $\ttbar + X$ cross sections, the PDF uncertainties are smallest for the NNPDF4.0 PDF set, followed by the ABMP16, MSHT20 and CT18 PDF sets.
Sizeable differences among the \chisq values are observed for the predictions obtained using different PDF sets, especially among those predictions which do not take into account the PDF uncertainties, providing a hint that these data are able to constrain uncertainties of many of the modern PDF sets. At first sight, the better agreement of the ABMP16 predictions with the experimental data could look surprising, given that this fit includes only $t\bar{t} + X$ total cross-section data, whereas the other fits also include some $t\bar{t} + X$ differential data. However, one should recall that the ABMP16 is the only one, among those fits, where PDFs are fitted simultaneously with $\alpha_s(M_Z)$ and $m_t$, whereas the other PDFs are fitted for fixed $m_t$ and $\alpha_s(M_Z)$ values.
In the recent analysis assessing the impact of top-quark production at the LHC on global analyses of NNPDF4.0 PDFs in Ref.~\cite{Kassabov:2023hbm}, it is claimed that this dataset cannot be reproduced by the NNLO SM predictions, resulting in a $22\,\sigma$ deviation, while here with the \chisq values we demonstrate that  the same dataset is perfectly consistent e.g.\ with the ABMP16 PDF set (with $\chisq/\text{dof}=20/34$ and the $p$-value of $0.97$) even using the nominal value $\mtpole=172.5$~GeV, and reasonably consistent with the NNPDF4.0 PDF set ($\chisq/\text{dof}=55/34$, $p=0.013$, which corresponds to $2.5\sigma$). 
The plots of the second row, all obtained with the ABMP16 PDFs, show that, the larger is the top-quark mass, varied in the range 170~GeV $< \mtpole < 175$~GeV, the smaller is the cross section for low \mtt close to the threshold, while the opposite is true for high $\mtt > 420$~GeV because of the cross-section normalization. 
From all panels of this row it is evident that the shape of the \ytt distribution is almost insensitive to the top-quark pole mass value. 
The plots of the third row show the behaviour of the distribution under
different ($\xi_R$, $\xi_F$) combinations. One can see that scale uncertainties increase at large \mtt, reaching up to ~$\pm$~3\% values in the highest \mtt bin, that are comparable to the data uncertainties in this kinematic region. Due to the cross-section normalization, the average size of the scale uncertainties for the normalized differential cross sections is a few times smaller than for the total cross section\footnote{When computing the normalized cross section, the same scale is used in the numerator and denominator.}. The largest cancellation of the scale uncertainties between numerator and denominator for the normalized differential cross sections happens at small values of \mtt and \ytt,
i.e. where scale uncertainties in the numerator and denominator have similar size, while for high values $\mtt \gtrsim 800$~GeV and $\ytt \gtrsim 1.5$ the size of the scale uncertainties for the normalized differential cross sections approaches the one for the total cross section. As for the case of the total \ttbar~$+~X$ cross sections, no uncertainty associated with the scale dependence of the cross section is included in the \chisq calculation, while later these uncertainties are propagated to the extracted values of \mtpole (see Section~\ref{sec:fit}).

Fig.~\ref{fig:top18004} presents a similar comparison, again for the \ytt distribution in different
\mtt bins, but this time obtained in the CMS study of Ref.~\cite{top18004} from the dileptonic decay channel of the top quarks.
In this experimental work, triple-differential distributions were considered, in $\ytt$, $\mtt$ and the number of additional jets $N_{j}$. Here we limit ourselves to double-differential distributions, because for a meaningful analysis of the $N_j$ distribution, merged predictions for $t\bar{t} + X$, $t\bar{t}j + X$, $t\bar{t}jj + X$, etc. should be considered. NNLO QCD predictions, however, at present are only available for $t\bar{t} + X$. In the first row of plots, results from different PDFs are compared, whereas in the second row, the effect of varying \mtpole at fixed PDF~+~$\alpha_s(M_Z)$ is shown.
The same trends as already noticed for the CMS data of Ref.~\cite{top20001} are observed even in this case, i.e., the shape of the \ytt distribution is better reproduced by the ABMP16 fit, than by other PDF sets, and the use of different \mtpole values does not have an impact on it, but only affects the normalization of the results. Theory predictions with all the considered central PDF sets slightly underestimate the data in the smallest \mtt bin, but are still compatible with them within 2$\sigma$. Like in Fig.~\ref{fig:top20001}, increasing the top-quark mass value decreases the value of the prediction of the \ytt distribution at small invariant masses, whereas at high \mtt, i.e. for \mtt $>$~400~GeV, the trend is the opposite due to the cross-section normalization.

\foreach \dataset in {top18004, a190807305, a200609274, top14013}{
\begin{figure}[h]
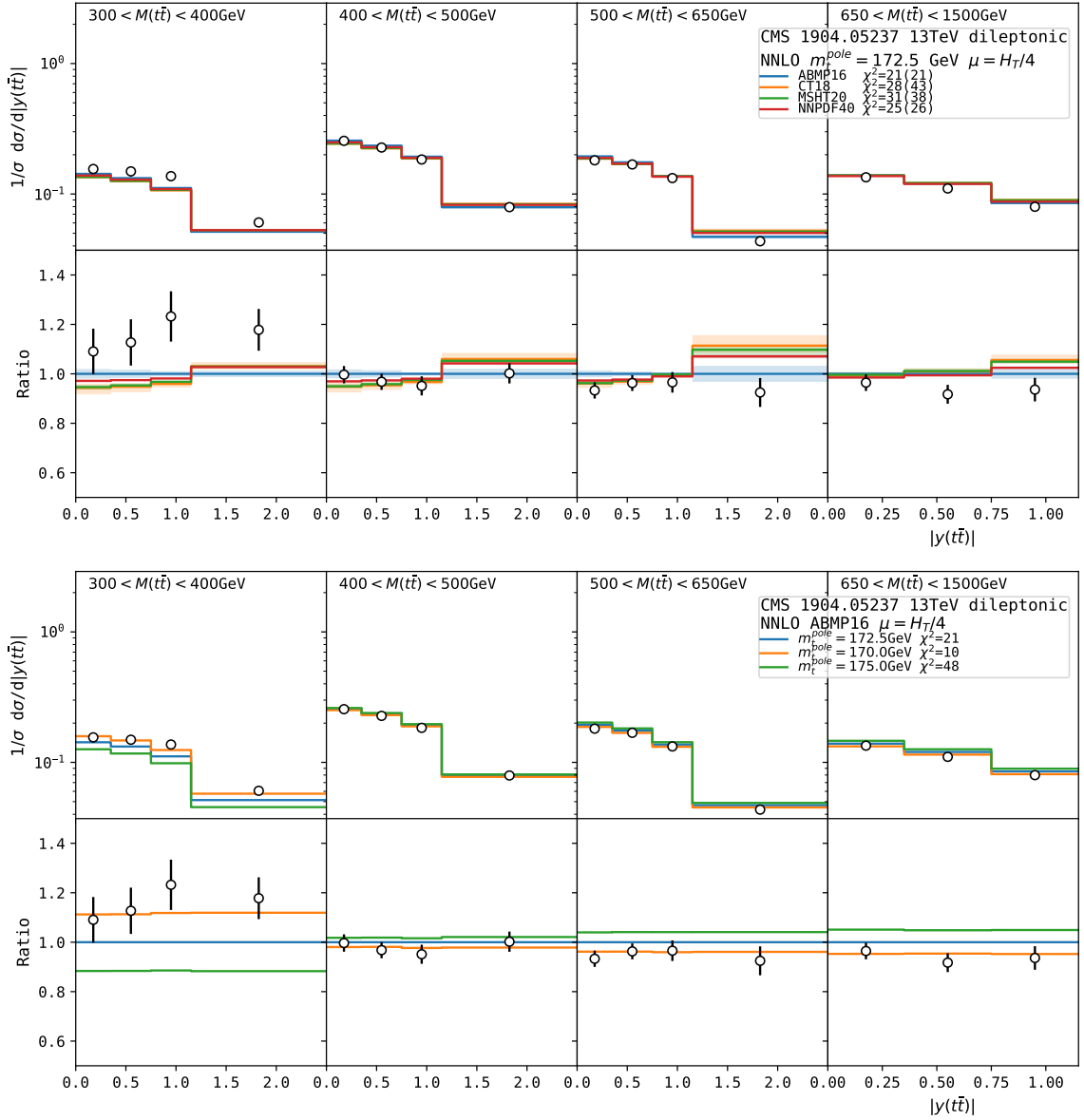

    \centering
    \includegraphics[width=1.00\textwidth]{figs/figs-data/pdfs_\dataset.pdf}
    \includegraphics[width=1.00\textwidth]{figs/figs-data/mt_\dataset.pdf}
    \caption{Comparison of the experimental data from Ref.~\cite{\dataset} to the NNLO predictions obtained using different PDF sets (upper) and, for the ABMP16 central PDF member, different \mtpole values (lower).}
    \label{fig:\dataset}
\end{figure}
}

Fig.~\ref{fig:a190807305} presents the \mtt distribution in different \ytt bins, corresponding to different panels. Theory predictions with different PDF~+~$\alpha_s(M_Z)$ sets at fixed \mtpole value (first row) and for different \mtpole values in case of the ABMP16 fit (second row), are compared to the ATLAS data of Ref.~\cite{a190807305} with the \ttbar-quark pairs decaying in the semileptonic channel. 
One can make similar comments as for Fig.~\ref{fig:top20001} and Fig.~\ref{fig:top18004}, i.e. that 
for small \mtt increasing the top-quark mass value decreases the predictions, vice versa in the high \mtt tails. Additionally, from the four panels in the first row, it is clear that the theory predictions agree with the data within larger experimental uncertainties even for small \mtt, differently from what has been observed in case of the CMS experimental data of Fig.~\ref{fig:top18004} and Fig.~\ref{fig:top14013} (see the plot in the first panel of the first row).
However, among all the \chisq values per degree of freedom for the theory and data comparison in Figs.~\ref{fig:top20001}--\ref{fig:a160707281_8tev} the best ones vary within $0.25 \lesssim \chisq/\text{dof} \lesssim 1$, and the corresponding $p$-values indicate that the experimental uncertainties might be conservative. 
It would be interesting to understand the origin of this difference, that might be related to technical details of the analyses performed independently by the two collaborations (ATLAS and CMS). 
In particular, to some extent experimentally measured cross sections always depend on the value of the top-quark mass parameter used in the Monte-Carlo (MC) simulations. The MC simulations are a necessary ingredient to unfold the detector-level cross sections to the parton level. In the CMS analysis of Ref.~\cite{top18004} such a dependence was studied, and a special analysis technique, called ``loose kinematic reconstruction'', was developed in order to avoid  the dependence on the MC top-quark mass. 
In that analysis, the loose kinematic reconstruction was used to measure triple-differential cross sections as a function of \mtt, \ytt, and extra jet multiplicity. 
As a result, for this measurement the dependence of the measured \ttbar~+~$X$ cross sections on the top-quark MC mass was demonstrated to be negligible (see Fig.~C.1 in Ref.~\cite{top18004} and discussion in Section~\ref{sec:data}). No such studies were reported in the papers describing the other CMS or ATLAS $\ttbar + X$ differential measurements.

Fig.~\ref{fig:a200609274} shows again the \ytt distribution in \mtt bins, this time for the ATLAS experimental data of Ref.~\cite{a200609274}, with the $t\bar{t}$-quark pairs decaying in the all-hadronic channel. Considering the larger data uncertainties, one can see a reasonable agreement between theory predictions and experimental data in all bins. The shape of distributions with different PDFs is in qualitative agreement with the one already observed in the case of the analyses of $t\bar{t}$ in the dileptonic and semileptonic decay channels. 
Theory predictions slightly overestimate the data in the two largest $\mtt$ bins at $\mtt>700$ GeV. Thus, this data set agrees better with theoretical predictions with a lower value of \mtpole,
however, due to the very wide $0<\mtt<700$~GeV bin, the sensitivity of these data to \mtpole is small.

Fig.~\ref{fig:top14013} presents also a comparison between theory predictions and experimental data similar to those presented in Figs.~\ref{fig:top20001},~\ref{fig:top18004} and~\ref{fig:a200609274}, but considering the double-differential data in \ytt, \mtt of Ref.~\cite{top14013}, from an analysis of the CMS collaboration using the dileptonic \ttbar decay channel at $\sqrt{s} = 8$~TeV (Run~1). From the comparison with previous figures, all related to Run~2, one can observe again compatible trends, with the ABMP16 PDFs well reproducing the shape and the normalization of the $\ytt$ distribution for $\mtt >$~400~GeV, whereas the theory/data agreement for $\mtt <$~400~GeV is slightly worse, but still compatible within 2$\sigma$.  

Figs.~\ref{fig:a14070371}--\ref{fig:a160707281_8tev} refer to single-differential \mtt distributions, compared to Run 1 experimental ATLAS data at $\sqrt{s}=7$ and $8$~TeV. 
Considering the reduced amount of information with respect to the case of double-differential distributions, due to the integration over $y(t\bar{t})$ on the full phase-space for each \mtt bin and the small number of data points, all the measurements agree well with the theoretical predictions and have a limited sensi\-ti\-vi\-ty to PDFs and \mtpole. 
For each of these datasets, the best \chisq value is $\chisq < 2$ for $4\text{--}6$ degree of freedom, suggesting that even for these ATLAS measurements of \ttbar~+~$X$ differential cross sections performed during Run~1 the experimental uncertainties might be too conservative.

\foreach \dataset/\citation in {a14070371/\cite{a14070371}, a151104716/\cite{a151104716}, a160707281_7tev/\cite{a160707281} ($\sqrt{s}$~=~7~TeV), a160707281_8tev/\cite{a160707281} ($\sqrt{s}$~=~8~TeV)}{
\begin{figure}
    \centering
    \includegraphics[width=0.49\textwidth]{figs/figs-data/pdfs_\dataset.pdf}
    \includegraphics[width=0.49\textwidth]{figs/figs-data/mt_\dataset.pdf}
    \caption{Comparison of the experimental data from Ref.~\citation\xspace to the NNLO predictions obtained using different PDF sets (left) and, for the ABMP16 central PDF member, different \mtpole values (right).}
    \label{fig:\dataset}
\end{figure}
}

Considering all together the comparisons between the experimental data and theory predictions, both for the total and differential inclusive \ttbar~+~$X$ cross sections, we report an overall good agreement between the NNLO QCD predictions and the data. 
For all datasets, the best value of \chisq per degree of freedom does not exceed $\approx 1$, whereas for several ATLAS datasets this value and the corresponding $p$-value hint towards a possible overestimation of the experimental uncertainties. For the differential measurements, one can conclude that, apart from a few (\ytt, \mtt) bins, the experimental data from different analyses can be considered as roughly compatible among each other.

\section{NNLO fits of the top-quark pole mass value} 
\label{sec:fit}

We now focus on the fits of the top-quark pole mass value based on the NNLO description discussed in the previous section. The fits were performed using the \texttt{xFitter} framework~\cite{Alekhin:2014irh}, an open source QCD fit framework, initially developed for extracting PDFs, then extended to also extract SM parameters correlated with PDFs (e.g. heavy-quark masses) and, more recently, to constrain couplings in the SM as an effective field theory (SMEFT)~\cite{Anataichuk:2023elk}. 
To study the sensitivity to PDFs and $\alpha_s(M_Z)$, each fit was carried out using as input different PDF sets with their associated $\alpha_s(M_z)$. 
The same state-of-the-art sets considered in the previous section
(ABMP16, CT18, MSHT20 and NNPDF4.0) were considered also here.
As we will see in the following, the extracted \mtpole values associated with different (PDF~+~$\alpha_s(M_Z)$) sets are well compatible among each other, which justifies the procedure. A more sophisticated procedure would have involved a simultaneous fit of $m_t$, PDFs and $\alpha_s(M_Z)$, in line with the procedure by Ref.~\cite{Alekhin:2017kpj} at NNLO and by, e.g., Refs.~\cite{Alekhin:2018pai, top18004, Garzelli:2020fmd} at NLO. 
However, Ref.~\cite{Alekhin:2017kpj} includes only $t\bar{t}$~+~$X$ total cross sections, whereas in this work we also consider single- and double-differential distributions, which would make a simultaneous fit of the three quantities at NNLO quite a major effort. Additionally, due to the differential cross sections being normalized, the correlation degree between $\alpha_s (M_Z)$ and $m_t$, that is very large when considering absolute total cross-section data, is significantly reduced.  
We also observe that, while the ABMP16 fit has incorporated data on the total $t\bar{t} + X$ production cross section at the LHC Run 1, the CT18, MSHT20 and NNPDF4.0 ones have also included some single- and double-differential distributions from Run~1 and~2. No triple-differential $t\bar{t} + X$ distributions from Ref.~\cite{top18004} have been incorporated in the standard released version of any of these PDF fits, at least so far. 

On the other hand, dedicated studies on the impact of top-quark production cross sections on various recent PDF fits have also been published. In particular, Ref.~\cite{Bailey:2019yze} refers to the work building upon the MMHT PDFs, Ref.~\cite{Kadir:2020yml} describes the impact of ATLAS and CMS single-differential data
at $\sqrt{s}~=~8$~TeV on the CTEQ-TEA fit, Ref.~\cite{Ablat:2023tiy} 
explores the impact of single-differential data at $\sqrt{s}=13$~TeV on the CT18 PDF fit, Ref.~\cite{Hou:2019gfw} shows the impact of including $t\bar{t}$-quark pair distributions in the CT14HERA2 PDFs, Ref.~\cite{Kassabov:2023hbm} assesses the impact of \ttbar production at the LHC on the NNPDF PDFs and on Wilson coefficients in the SMEFT, whereas Ref.~\cite{Cridge:2023ztj} describes efforts to constrain the top-quark mass within the global MSHT fit. 
Our work employs consistently NNLO predictions (instead of NLO ones rescaled by means of $K$-factors) and considers a wider set of more specific (i.e. double-differential) state-of-the-art  $t\bar{t}~+~X$  experimental data than most of the previous ones, and it considers simultaneously multiple modern PDF fits, aiming to provide a more comprehensive overview.

\begin{figure}[h!]
    \centering
    \includegraphics[width=0.49\textwidth]{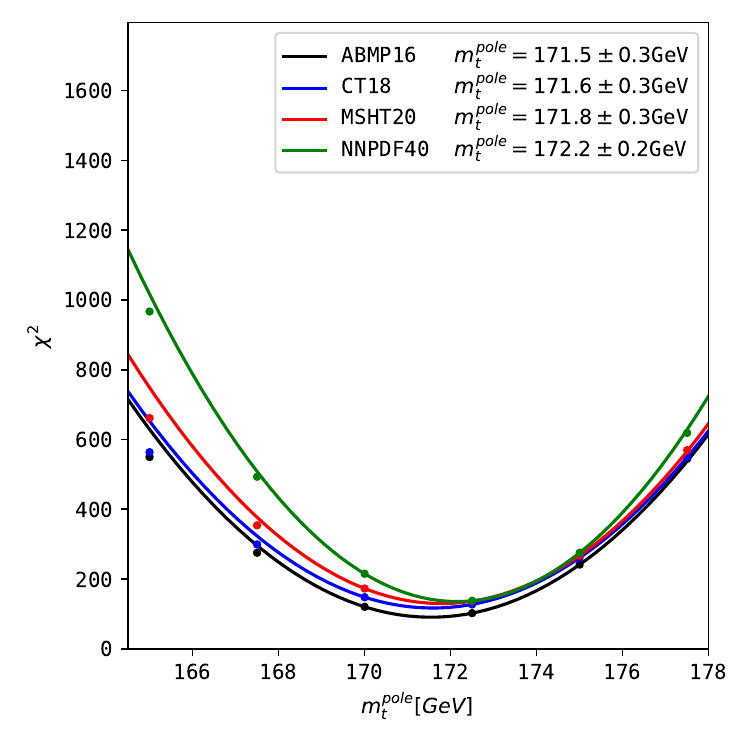}
    \includegraphics[width=0.49\textwidth]{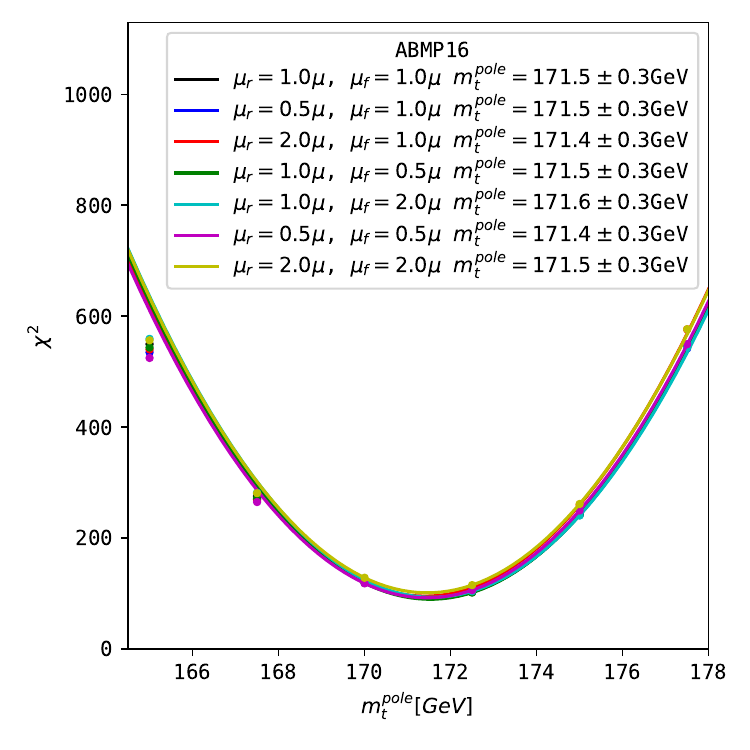}
    \caption{The \mtpole extraction at NNLO from all experimental data using different PDF sets (left) and using the ABMP16 PDF set with different scale choices (right).}
    \label{fig:chi2}
\end{figure}

In each fit, the value of \mtpole is extracted by calculating a \chisq from data and correspon\-ding theoretical predictions for a few values of \mtpole, and approximating the dependence of the \chisq on \mtpole with a parabola. The minimum of the parabola is taken as the extracted \mtpole value, while the uncertainty on the latter is derived from the $\Delta\chi^2 = 1$ variation. In this prescription one assumes a linear dependence of the predicted cross sections on \mtpole.
As an example, in Fig.~\ref{fig:chi2} the \chisq profiles for the \mtpole scans are shown in case of our most global fit, including total and differential cross-section data from both Run~1 and Run~2. The left panel shows that the profiles obtained using as input different PDF sets all have a parabolic shape with a clear minimum and are quite similar among each other.
We use the values $\mtpole=170$, $172.5$ and $175$~GeV to build each parabola, but we also show the \chisq values obtained repeating the computations with an input of $\mtpole=165, 167.5$ and $177.5$~GeV. The latter values agree with the parabola reasonably well, thus justifying the linear dependence assumption, even outside the range of $170<\mtpole<175$~GeV. 
The top-quark mass value at the minimum value of $\chisq$
varies by up to 0.6~GeV between the PDF sets considered.
For the ABMP16 PDF set, the value of \chisq at the minimum is $\chisq=101$ for 121 dof (corresponding to the total of 122 data points). 
From the right panel, obtained using as a basis the ABMP16 PDFs, it is clear that the extracted top-quark mass value is quite stable with respect to the ($\mu_R$, $\mu_F$) 7-point scale variation by a factor of two around the central ($\mu_R$, $\mu_F$) scale. For this variation, the difference in the central value of the extracted top-quark mass amounts to a maximum of 0.2~GeV. On the other hand, the uncertainty on \mtpole derived from a $\Delta\chisq$ = 1 variation, that accounts for the uncertainties of ABMP16 PDFs, the data and other sources included in the covariance matrix, see eq.~(\ref{eq:covmat}), and thus excludes scale uncertainties, amounts to $0.3$~GeV.  
We assign the maximum difference on the $\chi^2$ values from the ($\mu_R$, $\mu_F$) 7-point scale variation as a scale variation uncertainty on the extracted \mtpole value. When treated in this way, the scale variation uncertainty cannot be constrained by data (see also the related comment in Section~\ref{sec:prefit}). 
These uncertainties are asymmetric, as a consequence of the fact that the changes of the cross sections under scale variations are asymmetric.
The $\chisq$ values obtained in the fit for each data set are summarized in Table~\ref{tab:data_chi2}.

\begin{table}[htb]
\setlength{\tabcolsep}{6pt}
\centering
\begin{tabular}{llllll}
Data set & $n$ & ABMP16 & CT18 & MSHT20 & NNPDF4.0 \\
\hline
CMS 13 TeV semileptonic \cite{top20001} & 34 & 19(20) & 29(176) & 38(132) & 55(90) \\
CMS 13 TeV dileptonic \cite{top18004} & 15 & 15(15) & 23(38) & 27(34) & 23(23) \\
ATLAS13 TeV semileptonic \cite{a190807305} & 19 & 11(15) & 12(17) & 11(13) & 12(12) \\
ATLAS 13 TeV all-hadronic \cite{a200609274} & 10 & 11(11) & 16(19) & 16(17) & 14(14) \\
CMS 8 TeV dileptonic \cite{top14013} & 15 & 11(15) & 11(12) & 11(12) & 12(12) \\
ATLAS 8 TeV semileptonic \cite{a151104716} & 6 & 10(12) & 4(4) & 4(4) & 5(5) \\
ATLAS 7 TeV dileptonic \cite{a160707281} & 4 & 2(3) & 1.9(1.9) & 1.6(1.6) & 1.1(1.1) \\
ATLAS 8 TeV dileptonic \cite{a160707281} & 5 & 0.2(0.2) & 0.4(0.5) & 0.4(0.4) & 0.2(0.2) \\
ATLAS 7 TeV semileptonic \cite{a14070371} & 4 & 0.9(1.0) & 5(6) & 6(6) & 3(3) \\
$\sigma(\ttbar)$ \cite{tca78,ta5,tc5,top20001,tc13ll,ta13lj,ta13ll,ta136,tc136} & 10 & 11(26) & 16(61) & 16(43) & 11(12) \\
\hline
Total  & 122 & 101(117) & 115(337) & 113(262) & 129(172) \\
\end{tabular}
\caption{The global and partial $\chisq$ values for each data set with its number of data points ($n$) obtained in the \mtpole extraction using different PDF sets and the best-fit values of \mtpole. An additional \chisq value is indicated in parentheses, which omits the PDF uncertainties.}
\label{tab:data_chi2}
\end{table}

\begin{figure}[htb]
    \centering
    \includegraphics[width=0.49\textwidth]{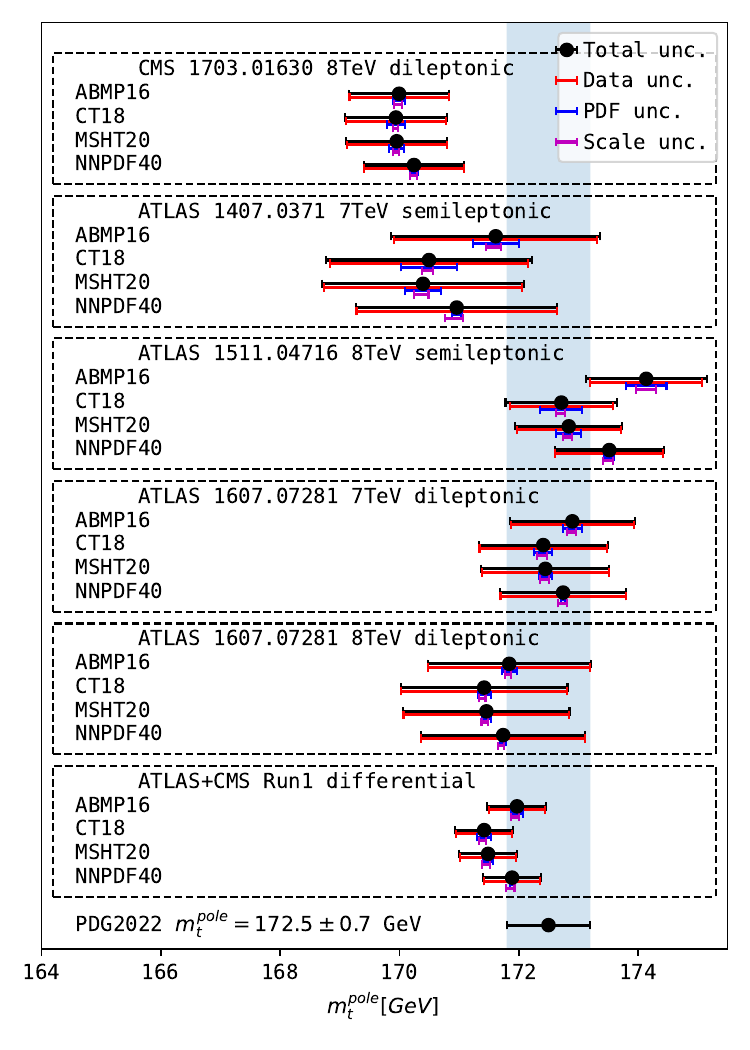}
    \includegraphics[width=0.49\textwidth]{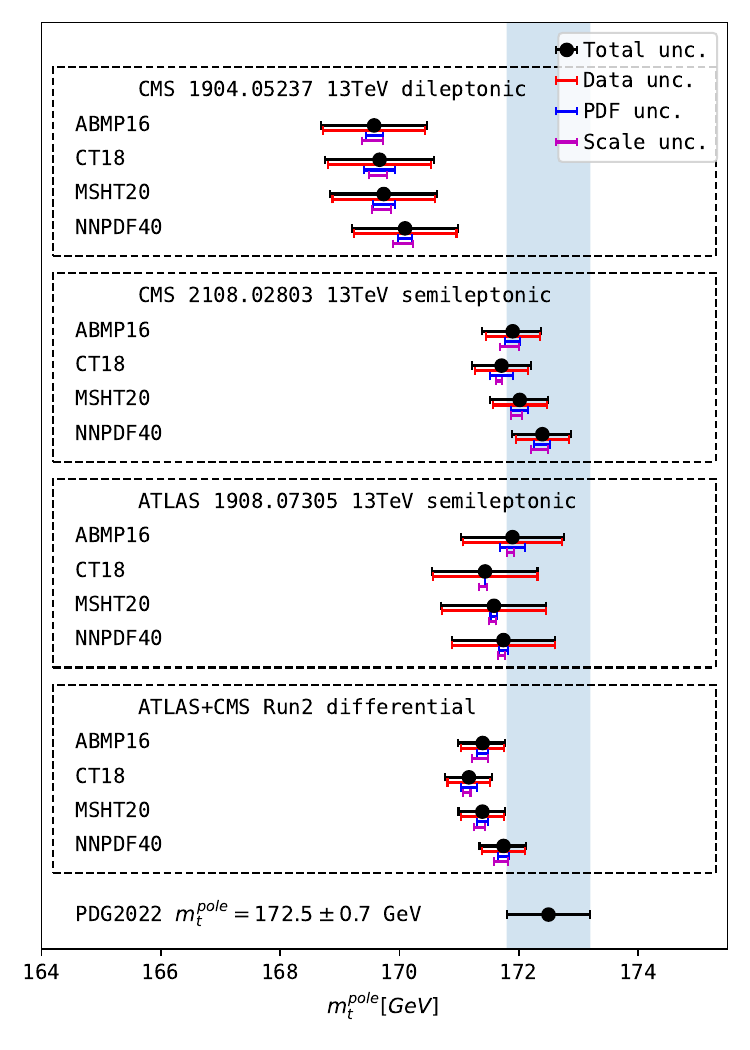}
    \caption{The \mtpole values extracted at NNLO from Run1 (left) and Run2 (right) measurements of differential $\ttbar+X$ cross sections.
    }
    \label{fig:mt-run12}
\end{figure}

\begin{figure}[htb]
    \centering
    \includegraphics[width=0.92\textwidth]{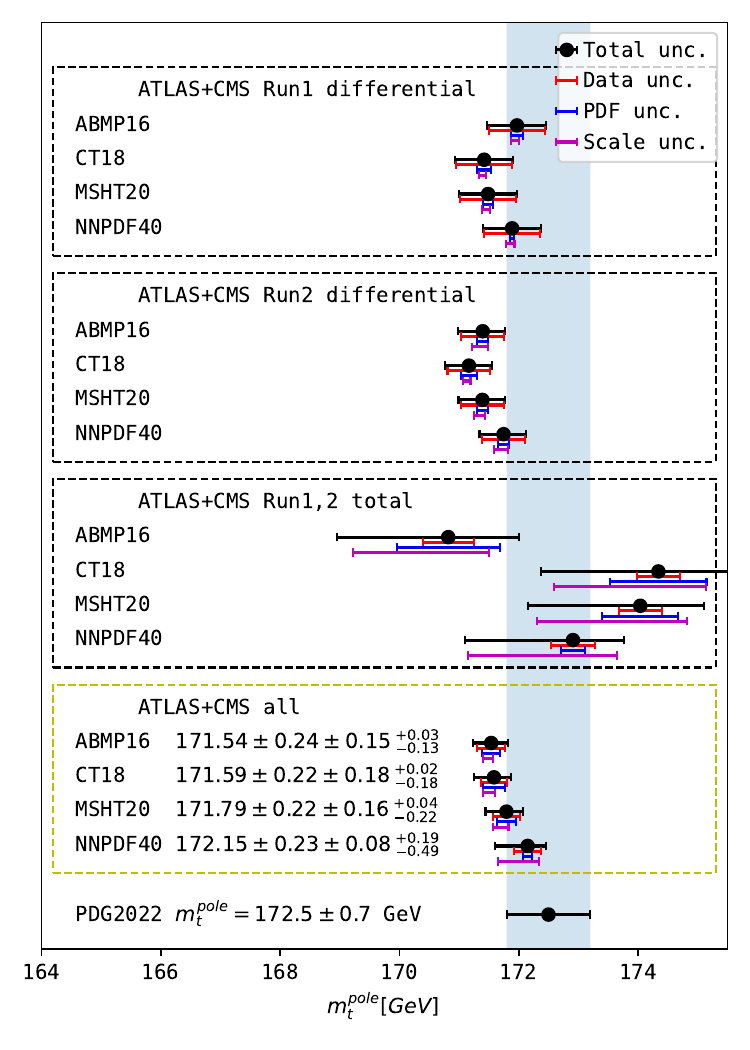}
    \caption{Summary of the \mtpole values extracted from Run1 and Run2 measurements of differential and total \ttbar $+ X$ cross sections
(the first two insets are the same as the lowest insets in Fig.~\ref{fig:mt-run12}.).
    }
    \label{fig:mt}
\end{figure}

\begin{figure}[htb]
    \centering
    \includegraphics[width=0.49\textwidth]{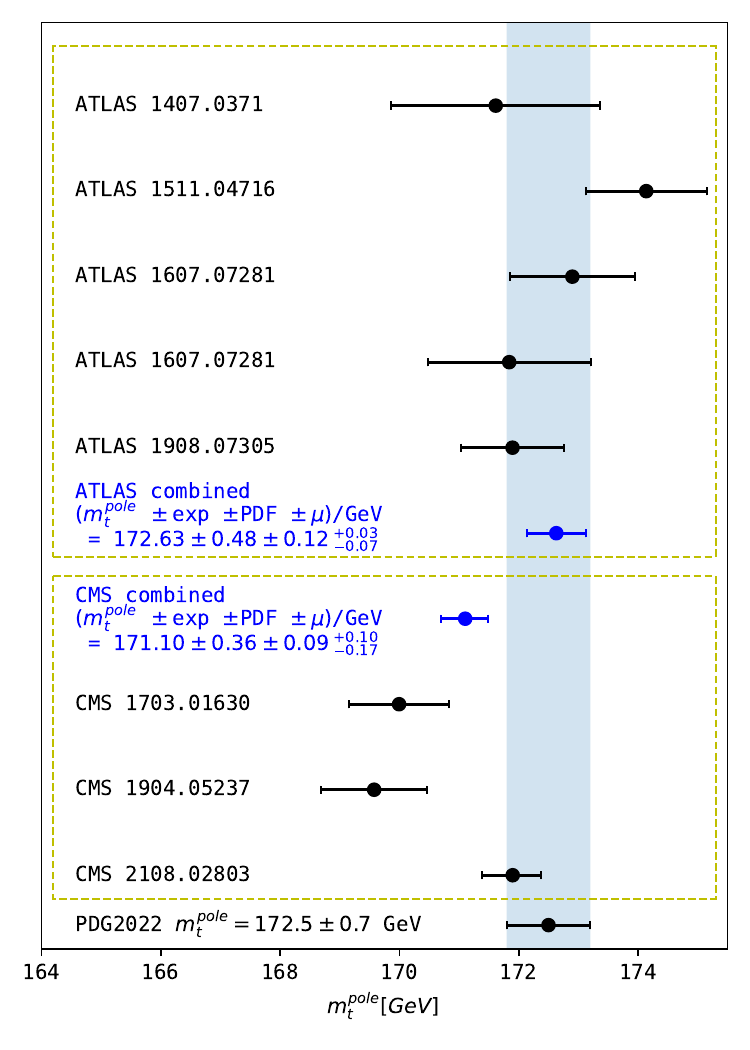}
    \includegraphics[width=0.49\textwidth]{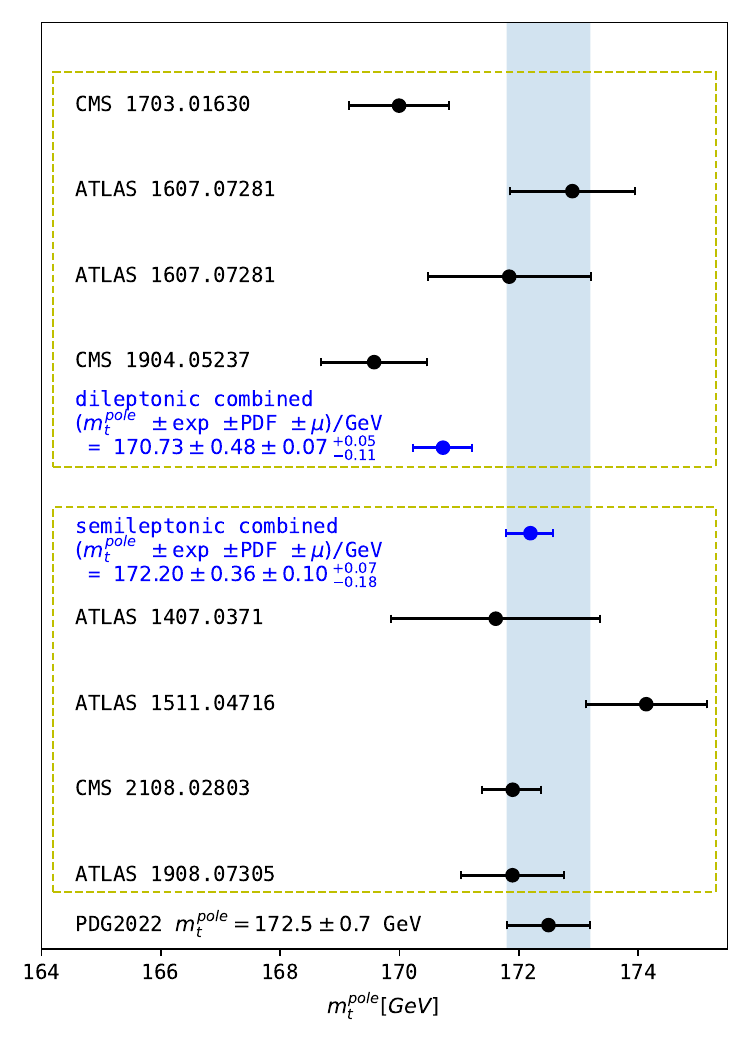}
    \caption{The \mtpole values extracted at NNLO using the ABMP16 PDF set from ATLAS and CMS measurements (left), and from dileptonic and semileptonic \ttbar decay channels (right).
      }
    \label{fig:mt-exp}
\end{figure}

In Figs.~\ref{fig:mt-run12} and \ref{fig:mt} the results for the \mtpole extraction from various (groups of) ex\-pe\-ri\-men\-tal datasets are shown.
In the left and right upper panels of Fig.~\ref{fig:mt-run12}, results related to Run~1 and Run~2 measurements of differential \ttbar~+~$X$ cross sections are shown separately, considering the fit to each individual Run 1 (Run 2) dataset, and the fit to all Run 1 (Run 2) differential datasets. 
Figure~\ref{fig:mt} reports the results of the fit to the Run~1~+~Run~2 data on total cross sections only, as well as the results of the global fit, including both total and single/double-differential cross-section Run 1 + Run 2 data.
In the right panel of Fig.~\ref{fig:mt-run12} we do not show as a separate plot the results of the fit to the ATLAS measurement in the all-hadronic \ttbar decay channel of Ref.~\cite{a200609274}, since it has a very low sensitivity to \mtpole, and the \mtpole uncertainties amount to $\sim 5$ GeV when using this dataset alone, but this measurement is in any case considered in both the Run 2 and (Run 1 + Run 2) global fits.
By comparing the two panels in Fig.~\ref{fig:mt-run12}, one can see that for fits to individual datasets the fit uncertainty component due to data uncertainties~\footnote{For brevity, the impact of the small numerical theoretical uncertainty of 1\% (which would be barely visible) was added to the data uncertainties on these plots.} is often larger in the case of Run 1 than Run 2 datasets. 
In the Run 2 analyses, in general, more accurate systematic uncertainty estimates have taken place, sometimes leading to an increase of the latter with respect to Run 1 cases. On the contrary, the statistics uncertainties in the Run~2 data are indeed smaller. The balance between these two trends reduces data-related uncertainties by a factor $\sim$ 1.5.
On the other hand,  data uncertainties in the case of fits to $t\bar{t}$ data in the dileptonic channel can be larger or smaller than those in the semileptonic channel, depending on the details of the analysis. In the dileptonic channels, the neutrino and antineutrino are not detected, and their tri-momenta (six unknowns) are reconstructed on the basis of kinematic considerations, whereas in the semileptonic channel only one of them needs to be reconstructed, which is an advantage when reconstructing differential distributions. However, jet energy scale uncertainties may become important in the semileptonic channel, due to the presence of two light jets (absent in the dileptonic channel). The dileptonic channel is less sensitive to pile-up, and when studying the ($e^\pm$, $\mu^\mp$) signature, becomes the most practical one for measuring total cross sections, due to the absence of Drell-Yan background. 
For more information on the peculiarities of each analysis and the kinematic reconstruction techniques, the reader can refer to the corresponding experimental paper (see the lists in Tables~\ref{tab:data_total}  and~\ref{tab:data_diff}). 
Both in case of Run~1 and in case of Run~2 global fits, the \mtpole central values obtained for each PDF set are compatible among each other, and, although systematically slightly lower, also compatible with the value $\mtpole=172.5\pm 0.7$~GeV reported in the PDG~\cite{ParticleDataGroup:2022pth}, when considering the uncertainties. 
The results of the extraction using either differential or total \ttbar~+~$X$ cross sections agree with each other within $\approx 1\sigma$, for any PDF set. 
We consider the compatibility of the results obtained as a sign of their robustness.
The global fit to the ensemble of Run 2 datasets shows slightly smaller data-related uncertainties than the global fit to the ensemble of Run 1 datasets, whereas scale uncertainties are similar (and slightly larger or smaller, depending on the PDF) in the two cases. 
One can also observe that data related to $t\bar{t}$ decays in the dileptonic channel (Refs.~\cite{top18004,top14013}) point towards central \mtpole values smaller than data related to decays in the semileptonic channel (Ref.~\cite{a190807305} by ATLAS and Ref.~\cite{top20001} by CMS). 
To explore this further, in Fig.~\ref{fig:mt-exp} we compare the values of \mtpole extracted using the ABMP16 PDF set from differential measurements which are grouped either by the experiment or decay channel.
The values extracted from all ATLAS and all CMS differential measurements are compatible within $2.5\,\sigma$, and the same level of compatibility is observed for the results extracted from the measurements in the dileptonic or semileptonic \ttbar decay channels.
In both cases, the difference originates almost entirely from the CMS measurements of Refs.~\cite{top14013,top18004} which point to a lower value of \mtpole than all the other measurements.

As for the theoretical uncertainties on the extracted \mtpole values, noticeably, the PDF uncertainties are larger for single-differential measurements than for the double-differential ones. 
We attribute this to the fact that the usage of the \ytt observable provides an added value to constrain the PDF uncertainties and decorrelate the gluon PDF and \mtpole. 
In general, all extractions from the differential cross-section measurements are dominated by the data uncertainties, while the PDF and scale variation uncertainties are a few times smaller. 
Even for the combined extraction from Run 1 or 2 differential cross-section measurements, both the PDF and scale uncertainties are about a factor two smaller than the data uncertainties. 
This is very different from the behaviour of the fits to the absolute total \ttbar cross sections: in this case, the dominant uncertainties come from the scale variations, followed by the PDF uncertainties and then data uncertainties for all PDF sets, except NNPDF4.0. 
For the latter, the PDF uncertainties are even smaller than the data uncertainties (see also Fig.~\ref{fig:total}). 
Also, for the case of the absolute total \ttbar~+~$X$ cross sections the central values of \mtpole obtained using different PDF sets differ by up to $\approx 3$~GeV. 
Part of this difference should be attributed to the fact that different PDF groups use different values of $\alpha_s(M_Z)$ and \mtpole in their fits.
We also observe that the biggest impact on the top-quark mass values extracted in our work comes from the analysis of double-differential normalized cross sections. In the normalization, the dependence on $\alpha_s$ largely cancels.

Let us look closer at the results of extraction of \mtpole from the total and differential cross sections obtained using different PDF sets.
Due to the external treatment of scale  variations (as explained above, the corresponding uncertainty is not included in the \chisq values, but is calculated as the difference of the nominal result and the ones with varied scales), the corresponding uncertainty is not reduced when combining e.g.\ the total and differential results. 
This leads to the situation that the scale variation uncertainty obtained from the global set of total and differential measurements is larger than the one obtained from the differential measurements only. 
This is most prominent when using the NNPDF4.0 set, because it has the smallest PDF uncertainties. Therefore in the extraction using the total and differential cross sections the total cross-section measurements receive a larger weight than when using other PDF sets, and the combined result appears to be more significantly affected by the scale variation uncertainty arising from the total cross sections.
Contrary, for all PDF sets the data and (to some extent) PDF uncertainties are reduced in the combined \mtpole extraction.

In principle, under such a treatment of the scale variation uncertainties one might want to even refrain from using the absolute cross sections together with the differential ones for the \mtpole extraction, in order to avoid to enhance the scale variation uncertainties on the final result. 
However, in the present situation we see that the resulting scale variation uncertainties when using all the measurements do not increase significantly. In particular, when using ABMP16, CT18 or MSHT20 sets, the scale variation uncertainties on the final result remain smaller that the data uncertainties, while they are slightly larger when using the NNPDF4.0 set due to the reasons discussed above. 
Furthermore, we notice that the scale variation uncertainties on the final \mtpole value in the case of CT18 or MSHT20 are slightly larger than in the case of ABMP16. 
This is also due to the different treatment of the scale variation uncertainties: indeed, from the lower panel of Fig.~\ref{fig:mt} one can see that the \mtpole values extracted from the differential and total cross sections agree better when using the ABMP16 PDF set (within $\approx 1$~GeV) than the CT18 or MSHT20 sets (within $\approx 2\text{--}3$~GeV). 

In summary, adding the differential data to the fit only including total cross sections plays a crucial role in decreasing the uncertainties on \mtpole by a factor of $\sim$ 3. The result of the most comprehensive fit has an uncertainty band ranging from 0.3 to 0.5~GeV, depending on the PDF set (due to the fact that scale and PDF uncertainties depend on PDFs), e.g.\ using the ABMP16 PDF set we obtain:
\begin{equation}
\label{eq:mtpole}
    \mtpole = 171.54 \pm 0.24\,\textrm{(exp)} \pm 0.15\,\textrm{(PDF)} {}^{+0.03}_{-0.13}\,{(\mu)}~\textrm{GeV} = 171.54 {}^{+0.28}_{-0.31}~\textrm{GeV},
\end{equation}
while the values obtained using the other PDF sets are provided in Fig.~\ref{fig:mt}.
These uncertainties are factor 2.5 smaller when compared to those in the most recent average $\mtpole=172.5\pm 0.7$~GeV of the PDG~\cite{ParticleDataGroup:2022pth}. They are similar to the results of Ref.~\cite{Cridge:2023ztj}~\footnote{In Ref.~\cite{Cridge:2023ztj} the value $\mtpole=173.0\pm 0.6$~GeV is obtained using a dynamic tolerance approach, while also the authors quote the value $\mtpole=173.0\pm 0.3$~GeV which would be obtained with a $\Delta\chisq = 1$ criterion. However, in this analysis scale variation uncertainties were not considered.}. 
One should also mention the existence of a renormalon ambiguity~\cite{Smith:1996xz} affecting \mtpole (but absent in short-distance mass definitions), leading to an intrinsic theoretical uncertainty on \mtpole in the range of $110$--$250$ GeV, corresponding to $\mathcal{O}(\Lambda_{QCD})$, see e.g. Ref.~\cite{Beneke:2016cbu,Hoang:2017btd}. This is not included in the uncertainty quoted in eq.~(\ref{eq:mtpole}) and affects all \mtpole determinations.

One should also observe that uncertainties related to the data used have similar size to the scale~+~PDF variation uncertainty for a fixed PDF set. We expect that forthcoming experimental data from Run~3 and Run~4 at the LHC, improving the statistical accuracy of the measurements, will lead to reduced data uncertainties on the extracted top-quark mass values. This will challenge theoretical capabilities of reducing theory uncertainties to at least a similar level as well.  

\section{Conclusions}
\label{sec:conclu}

Using the ATLAS and CMS measurements of absolute total and normalized single-dif\-fe\-ren\-tial and double-differential cross sections for $pp \rightarrow$ \ttbar $+ X$ production, compared to theoretical computations obtained with the \texttt{MATRIX} framework, we have extracted the top-quark pole mass \mtpole value at NNLO QCD accuracy. 
To do our fits, we have interfaced  the \texttt{MATRIX} framework to the \texttt{PineAPPL} library for the generation of interpolation grids, which can be convoluted very efficiently a-posteriori with any PDF + $\alpha_s(M_Z)$ set. The procedure allows for genuine NNLO predictions and fit results, without the use of any $K$-factor or approximation for relating NNLO predictions to lower-order ones. These approximations have indeed been adopted in various works for PDF and/or top-quark mass fitting via top-quark data, preceeding our one. In comparison to many previous works, we also use more specialized state-of-the-art data, in particular double-differential cross sections from a number of analyses, some of which have never been considered before for a top-quark mass extraction, to the best of our knowledge.
We used several state-of-the-art PDF~+~$\alpha_s(M_Z)$ sets as input. 
For the \mtpole extraction, we have propagated their PDF and/or $\alpha_s(M_Z)$ uncertainties, as well as the experimental data uncertainties (with correlations, where available). We have also estimated the uncertainties on \mtpole arising from renormalization and factorization scale variation, taking into account that their distribution is not Gaussian. In the global fit, the data uncertainties turn out to be of similar size as the combined scale and PDF uncertainty.
We have found that the fitted values of the top-quark mass using different PDF + $\alpha_s(M_Z)$ sets as input agree among each other within 1$\sigma$ uncertainty, with the best description of experimental data provided by the ABMP16 PDFs.
The extracted \mtpole values are compatible with the PDG 2022 average value and are accompanied by uncertainties smaller by a factor of roughly 2.5.

Data from Run 2, in particular those on normalized double-differential cross sections in Refs.~\cite{top20001,top18004,a190807305}, play a stronger constraining role with respect to the single-differential ones from Run 1. On the other hand, data on total inclusive cross sections collected in Run~1, 2 and 3 turn out to play only a minor constraining role with respect to the previous ones.
Upon combining together all differential experimental data from Run 1 and Run 2, we get consistent \mtpole values, and this occurs for each considered PDF + $\alpha_s(M_Z)$ set. Overall, the datasets are well described by the theoretical predictions, and the extracted \mtpole values using as input different PDF + $\alpha_s(M_Z)$ sets are all compatible among each other. This is a sign of the overall robustness of our conclusions,   
However, we have identified and discussed also tensions between individual data sets. In particular, experimental data which are collected using dileptonic and semileptonic $t\bar{t}$ decay channels point towards \mtpole values which are in a tension among each other, with \mtpole values from the analysis of dileptonic data smaller by roughly 1.5~GeV than those from the analysis of semileptonic ones, but still compatible within $2.5\, \sigma$ accuracy. We believe that these tensions require further investigation on the experimental side.

Our present work can be regarded as a proof-of-principle that 
a simultaneous fit of the top-quark mass value, PDFs and $\alpha_s(M_Z)$
at NNLO accuracy, considering the correlations among these quantities and
using state-of-the art total and multi-differential $\ttbar +X$ production
data, is within reach. We plan to perform such a fit in future work, 
upgrading the precision and accuracy of the NLO fit results we have presented
in Ref.~\cite{Garzelli:2020fmd}. 

\acknowledgments

We acknowledge the use of the BIRD cluster at DESY, where part of the computations were performed. 
We thank Sergey Alekhin for useful discussions and Daniel Britzger for feedback on the manuscript.
The work of  M.V.G. and S.~M. has been supported in part by the Bundesministerium f\"ur Bildung und Forschung under contract 05H21GUCCA.
The work of O.~Z. has been supported by the {\it Philipp Schwartz Initiative} of the Alexander von Humboldt foundation. 
M.V.G., S.M. and O.Z. are grateful to the Galileo Galilei Institute in Florence for hospitality and support during the scientiﬁc program on Theory Challenges in the Precision Era of the Large Hadron Collider, where part of this work was done.

\bibliographystyle{JHEP} 
\bibliography{ttnnlo}

\end{document}